\documentclass[12pt]{iopart}

\usepackage{graphicx}
\begin{document}

\title[Study of magnetoelectrics using the praphase concept and exchange symmetry]
      {Interpretation of magnetoelectric phase states using the praphase concept and exchange symmetry}

\author{N~V~Ter-Oganessian and V~P~Sakhnenko}

\address{Institute of Physics, Southern Federal University, 194 Stachki pr., Rostov-on-Don, 344090 Russia}

\ead{nikita.teroganessian@gmail.com}

\begin{abstract}
The majority of magnetoelectric crystals show complex temperature-magnetic field or temperature-pressure phase diagrams with alternating antiferromagnetic incommensurate, magnetoelectric, and commensurate phases. Such phase diagrams occur as a result of successive magnetic instabilities with respect to different order parameters, which usually transform according to different irreducible representations (IR) of the space group of the crystal. Therefore, in order to build a phenomenological theory of phase transitions in such magnetoelectrics one has to employ several order parameters and assume the proximity of various instabilities on the thermodynamic path. In this work we analyze the magnetoelectrics MnWO$_4$, CuO, NaFeSi$_2$O$_6$, NaFeGe$_2$O$_6$, Cu$_3$Nb$_2$O$_8$, $\alpha$-CaCr$_2$O$_4$, and FeTe$_2$O$_5$Br using the praphase concept and the symmetry of the exchange Hamiltonian. We find that in all the considered cases the appearing magnetic structures are described by IR's entering into a single exchange multiplet, whereas in the cases of MnWO$_4$ and CuO by a single IR of the space group of the praphase structure. Therefore, one can interpret the complex phase diagrams of magnetoelectrics as induced by a single IR either of the praphase or of the symmetry group of the exchange Hamiltonian. Detailed temperature-magnetic field phase diagrams of MnWO$_4$ and CuO for certain field directions are obtained and the magnetic structures of the field-induced phases are determined.
\end{abstract}

\pacs{75.85.+t, 77.84.-s}

\maketitle

\section{Introduction}

In the recent decade magnetoelectricity has become one of the
focal points of interest for both the magnetic and ferroelectric
communities. The potentialities of combining magnetic and
ferroelectric properties in the same system and, which is more
important, the possibility of controlling one property by altering
the other, open the way to new interesting applications. The
promising applications of the magnetoelectric effect include new
types of devices for reading, writing, and storage of information~\cite{Wang_Memory,Vopsaroiu_Memory}, various sensors~\cite{Wang_Sensor,Chen_Sensor}, microwave~\cite{Pyatakov_UFN_Review} and spintronics~\cite{Pantel_Spintronics} devices, and wireless energy transfer and energy harvesting technologies~\cite{Pyatakov_UFN_Review}.
Therefore considerable efforts are devoted from both the
theoretical and experimental points of view to the search of new
magnetoelectric crystals and explanation of their properties.

The vast experimental data accumulated during the past years
allows establishing specific features pertinent to
magnetoelectrics, which have been summarized in recent reviews.
The magnetoelectrics can be divided into two
classes~\cite{Khomskii_Review}. To the so-called type-I
magnetoelectrics, which are sometimes referred to as
ferroelectromagnets, belong crystals in which ferroelectricity and
magnetic order appear independently and have different sources.
The prominent examples are BiFeO$_3$~\cite{Catalan_Scott_Review}
and YMnO$_3$~\cite{WangLiuRen_Review}. Type-I multiferroics are
usually characterized by rather high ferroelectric transition
temperatures and large electric
polarization ($\sim$ 10 -- 100$~\mu\rm C/cm^2$), but the generally
large difference between the ferroelectric and magnetic transition
temperatures and the different causes of the two orders result in
small coupling between them~\cite{Scott_Blinc_Review}.

The second class, the type-II magnetoelectrics, is comprised of
multiferroics, in which electric polarization appears upon
magnetic phase transitions. This class is generally characterized
by much lower transition temperatures (10 -- 40~K) and rather low
electric polarization values (usually of the order of 10 --
100$~\mu\rm C/m^2$). The prominent examples of such
magnetoelectrics are rare-earth manganites RMnO$_3$ (R=Gd, Tb, and
Dy)~\cite{KimuraRMO3}. Compared to ferroelectromagnets the type-II
multiferroics provide much stronger magnetoelectric coupling due
to the fact that electric polarization is induced by magnetic
order. Therefore, application of external magnetic field alters
the magnetic structure, which may result in suppression,
induction, rotation, or a flop of electric polarization. In this paper we focus on the type-II magnetoelectrics.

Two main features of the type-II magnetoelectrics can be
established from the analysis of the vast experimental data
available to date. The first one consists in the fact that
magnetoelectrics often exhibit complex temperature--pressure or
temperature--magnetic field phase diagrams with alternating
magnetically ordered modulated, commensurate, and magnetoelectric
phases. Common appearance of incommensurate magnetically ordered
phases and the frequent emergence of electric polarization in them
determine the description of magnetoelectricity from both the
micro- and macroscopic points of view. Despite the fact that
ferroelectricity can be also induced by commensurate magnetism
(as, for example, in rare-earth manganates
RMn$_2$O$_5$~\cite{Kimura125manganites}), this manifests itself in
consideration of various complex magnetic orders such as screw,
cycloidal, helix, and others as sources of
magnetoelectricity~\cite{TokuraSeki_Review,Arima_Review}.

From the macroscopic crystal symmetry point of view the close
connection between the appearance of modulated and ferroelectric
phases in magnetoelectrics was recently pointed out by the present
authors~\cite{SakhnenkoImproperFerroelectric}. It was pointed out,
in particular, that in some magnetoelectrics the transformational
properties of magnetic order parameters are described by the
irreducible representations (IR) not satisfying the Lifshitz criterion.
This results in long-periodical modulation of the magnetic order,
as well as in ferroelectric states among the low symmetry phases
induced by these order parameters. Unlike other phenomenological models (see, for example,~\cite{Harris_Landau_analysis,Harris_Review_Order_parameters,Toledano_MnWO4}),
in which one employs separate order parameters for different
experimentally observed modulation wave vectors, our approach
provides the whole picture description using the minimal number of
order parameters. In the case when a commensurate magnetically
ordered phase is present in the phase diagram, the order parameters
belonging to its wave vector are chosen and the incommensurate
phases are found to be well described by the existing Lifshitz invariants. This is, for
example, the case of
RMn$_2$O$_5$~\cite{SakhnenkoImproperFerroelectric},
MnWO$_4$~\cite{SakhnenkoImproperFerroelectric,SakhnenkoMnWO4}, and
CuO~\cite{Sakhnenko_Praphase}. In other cases (such as, for
example, in CuCl$_2$~\cite{Sakhnenko_Exchange_FTT}), when the magnetic order does not lock-in to a commensurate structure, one may choose the commensurate wave vector
closest to that of the incommensurate phases to define the order parameters. This approach is justified by the fact that such models as, for example, the ANNNI model, which describe the magnetic structure by the Ising or Heisenberg Hamiltonians, show a multitude of incommensurate phases, but nevertheless usually have commensurate magnetic structure as the ground state~\cite{Selke_ANNNI,Hayden_Heisenberg}.

Another well established peculiarity of magnetoelectrics consists
in the following. Neutron diffraction studies of many
magnetoelectrics and the respective symmetry analysis reveal that
the description of the experimentally observed magnetic structures
requires two or more magnetic order parameters belonging to the
same wave
vector~\cite{Harris_Landau_analysis,Kenzelmann_Two_Order_Params,Lautenschlager_MnWO4,Toledano_CuO}.
These order parameters transform according to a single or different IR's of the symmetry group of the paraelectric and paramagnetic phase (paraphase) and induce electric polarization acting simultaneously. Indeed, in the case where the space group of the initial paraphase possesses spatial inversion and the resulting magnetic cell of the magnetically ordered phase coincides with the initial chemical cell (i.e., when $\vec{k}=0$) magnetic order parameters transforming according to a single IR cannot induce polarization~\cite{Kovalev_Kristallografiya}. However, this is generally not the case for $\vec{k}\neq0$, for which a single magnetic order parameter may induce electric polarization~\cite{SakhnenkoImproperFerroelectric}.

It is found that, for example, in
TbMnO$_3$~\cite{Kenzelmann_Two_Order_Params} and
MnWO$_4$~\cite{Lautenschlager_MnWO4} two different order parameters
condense successively, whereas in CuCl$_2$~\cite{Sakhnenko_Exchange_FTT,Seki_CuCl2} they condense
simultaneously. These order parameters transform according to
different irreducible representations of the crystal space group.
Therefore, for the correct phenomenological description of the
magnetic phase transitions in these magnetoelectrics it is
necessary to assume the proximity of instabilities with respect to
different irreducible representations on the thermodynamic
path~\cite{Toledano_MnWO4,Toledano_TbMnO3}.

In this paper we analyze several magnetoelectrics (MnWO$_4$, CuO, pyroxenes NaFeSi$_2$O$_6$ and NaFeGe$_2$O$_6$, Cu$_3$Nb$_2$O$_8$, $\alpha$-CaCr$_2$O$_4$, and FeTe$_2$O$_5$Br) from the point of view of crystal and exchange symmetry. We show that in all the considered cases their magnetic structures are described by IR's entering into a single exchange multiplet. For all the studied magnetoelectrics except for FeTe$_2$O$_5$Br we introduce praphase structures, which are more symmetric crystal structures that can be obtained by small atomic displacements. In the cases of MnWO$_4$ and CuO we show that their magnetically ordered phases are described by a single IR of the space group of the praphase. Therefore, the complex phase diagrams of the studied magnetoelectrics can be interpreted as induced by a single IR either of the space group of the praphase or of the symmetry group of the exchange Hamiltonian, which explains the proximity of various magnetic instabilities in magnetoelectrics. Furthermore, even when the introduction of the praphase does not lead to the reduction of the number of order parameters, taking into account latent pseudosymmetries allows determining the influence of external magnetic or electric fields or elastic stresses of certain symmetry on the phase transitions. For MnWO$_4$ and CuO we obtain temperature-magnetic field phase diagrams for certain field directions. The phase diagram for MnWO$_4$ is in excellent qualitative correspondence with the experimental one, whereas that for CuO predicts new magnetic field-induced phases. The magnetic structures of the new phases in CuO as well as those of the phases \textbf{HF}, \textbf{IV}, and \textbf{V} of MnWO$_4$ are discussed.

The paper is organized as follows. In \sref{sec:Exchange} we recall the basics of the symmetry of the exchange Hamiltonian and its relation to the crystal symmetry, in \sref{sec:Praphase} we briefly discuss the application of the praphase concept to the theory of phase transitions, in ~\sref{sec:PT_in_MEs} we perform the analysis of magnetic phase transitions in several magnetoelectrics using the exchange symmetry and the praphase concept, and in \sref{sec:Discussion} we discuss the obtained results.

\section{Exchange symmetry\label{sec:Exchange}}

Exchange interaction prevails in the magnetic structure formation of most of the magnetically ordered crystals~\cite{Izyumov_JMMM_4,Izyumov_Neutron_diffraction}. In the simplest case
the exchange Hamiltonian can be written as
\begin{equation}
H=-\sum_{nimj}J_{nimj}\left(\textbf{S}_{ni}\cdot\textbf{S}_{mj}\right),\label{EQ:Hamiltonian}
\end{equation}
where $\textbf{S}_{ni}$ is the $i$th atom spin operator in the
$n$th unit cell and $J_{nimj}$ is the exchange integral between
the atoms $ni$ and $mj$. In the system of spins exceeding $\frac{1}{2}$ the
Hamiltonian~\eref{EQ:Hamiltonian} contains terms of higher order
with respect to
$\left(\textbf{S}_{ni}\cdot\textbf{S}_{mj}\right)$. Many-spin
exchange is also possible.

The exchange Hamiltonian~\eref{EQ:Hamiltonian} depends only on
the mutual orientation of spins and is, therefore, invariant with
respect to simultaneous arbitrary angle rotation of all spins
around any axis. The orientation of the magnetic structure
relative to the crystal axes is determined by the relativistic
interactions (e. g. spin-orbit or dipole-dipole interaction),
which in many cases are weaker than the exchange interaction~\cite{Izyumov_JMMM_4,Izyumov_Neutron_diffraction}.

When building a phenomenological theory of magnetic phase
transitions in crystals one usually proceeds from the symmetry of
the disordered paramagnetic phase. The experimentally observed
magnetically ordered states are then classified according to IR's of space group of the paramagnetic
phase. Thus, the exact symmetry of the exchange as well as
relativistic interactions is taken into account. However, the
symmetry of exchange interactions is higher, than that of the
relativistic ones. Consequently, in cases when the exchange
interaction prevails in determining the magnetic structure, the
information on the exchange interaction symmetry is lost.
Additional symmetry of the exchange interactions leads to
additional exchange energy degeneracy of magnetic states, compared
to that corresponding to the space group symmetry~\cite{Izyumov_Neutron_diffraction}.
Therefore, one first has to consider the symmetry of the exchange
Hamiltonian and then take into account the exact space symmetry of
the crystal.

The symmetry group of the exchange Hamiltonian (the exchange
group) is given by the direct product $G_a\otimes O_s\otimes I_s$,
where $G_a$ is the space group acting on the atom coordinates,
$O_s$ is the rotational group in the spin space, and $I_s$ is the
inversion group containing the unit element and spin inversion~\cite{Izyumov_Neutron_diffraction}. It
can be shown that every IR $d^{\{\textbf{k}\}\nu}$ of the crystal space group, which is
characterized by the wave vector star $\{\textbf{k}\}$ and which enters
into the permutational representation of the atoms forming the
magnetic structure, induces IR $d^{\{\textbf{k}\}\nu}\otimes V'$ of the exchange group~\cite{Izyumov_Neutron_diffraction,Izyumov_JMMM_Exchange_Miltiplet}. Here $V'$
is the representation giving the transformational properties of a
pseudovector. To determine the connection between the irreducible
representation of the exchange group with those of the space group
one has to expand the limitation of $d^{\{\textbf{k}\}\nu}\otimes
V'$ on the space group with respect to its IR's~\cite{Izyumov_JMMM_Exchange_Miltiplet}
\begin{equation}
d^{\{\textbf{k}\}\nu}\otimes V'=\sum_\mu r_\mu^\nu
d^{\{\textbf{k}\}\mu}.\label{EQ:MultipletExpansion}
\end{equation}
Here $r_\mu^\nu$ are the expansion coefficients. The irreducible
representations entering into
expansion~\eref{EQ:MultipletExpansion} are degenerate with
respect to exchange energy and form the exchange multiplet~\cite{Izyumov_JMMM_Exchange_Miltiplet}. The
splitting of the energy of states corresponding to different IR's in the exchange multiplet is
determined by anisotropic relativistic interactions and in many cases is small
due to the smallness of these interactions compared to the
exchange energy. Therefore, the magnetic structure may be determined by a set of IR's of the space group, which form the exchange multiplet~\cite{Izyumov_JMMM_4,Izyumov_Neutron_diffraction,Izyumov_JMMM_Exchange_Miltiplet}.

\section{The praphase concept\label{sec:Praphase}}

The praphase concept is widely used in many fields of science from
elementary particle physics to the solid state. In the phase
transition theory in crystals the praphase concept is widely used, for example, in description of reconstructive phase transitions~\cite{Toledano_BOOK}. Adapting it to the needs of the present work the concept can be briefly described as follows. When building a phenomenological theory of phase
transitions one starts with a symmetry $G$ of the parent phase.
Sometimes the crystal structure of the parent phase can be
transformed by small relative displacements of the constituting
atoms to a structure of higher symmetry $G_{\rm p}$. One may expect that
this high symmetry structure, referred to as praphase, is realized
upon temperature increase, however the decomposition or melting of
the sample may occur earlier. The phase transition $G_{\rm p}\rightarrow
G$ is described by a generally multicomponent order parameter
$\{\eta_i\}$. Thus, one can base the phase transition model on the
higher symmetry praphase structure taking into account that
$\eta_i\neq0$.

The introduction of praphase improves the phase transition model
by taking into account latent pseudosymmetries in the object under
study. In the following we analyze a set of magnetoelectrics using
the praphase concept and exchange symmetry. The praphase structures of the pyroxenes and Cu$_3$Nb$_2$O$_8$ were found using the program PSEUDO located at the {\it Bilbao Crystallographic Server} (http://www.cryst.ehu.es/)~\cite{PSEUDO}.

\section{Phase transitions in magnetoelectrics\label{sec:PT_in_MEs}}

\subsection{MnWO$_4$\label{sec:MnWO4}}

Wolframite MnWO$_4$ has become one of the prominent examples of magnetoelectrics since it shows incommensurate paraelectric and ferroelectric magnetically ordered phases, low-temperature commensurate magnetic phase, and complex temperature -- magnetic field phase diagrams. At room temperature it possesses a monoclinic structure described by the space group $P2/c$ (C$_{2h}^4$)~\cite{Lautenschlager_MnWO4}. On lowering the temperature MnWO$_4$ undergoes a sequence of magnetic phase transitions at $T_{\rm N}$=13.5~K, $T_2$=12.7~K, and $T_1$=7.6~K, which lead to the appearance of magnetically ordered states \textbf{AF3}, \textbf{AF2}, and \textbf{AF1}, respectively~\cite{Taniguchi06_MWO}. The structure of the low-temperature commensurate magnetic phase \textbf{AF1} is described by the wave vector $\vec{k}_{\rm c}=(\frac{1}{4},\frac{1}{2},\frac{1}{2})$, whereas the incommensurate phases \textbf{AF2} and  \textbf{AF3} are characterized by the wave vector $\vec{k}_{inc}=(-0.214,\frac{1}{2},0.457)$~\cite{Lautenschlager_MnWO4}. In the phases \textbf{AF1} and \textbf{AF3} the collinearly aligned magnetic moments are confined to the $ac$ plane forming an angle of about $35^\circ$ with the $a$ axis (this direction is hereafter referred to as the easy axis), whereas in the \textbf{AF2} phase there appears an additional component along the $b$ axis.

Electric polarization along the $b$ axis appears at the second phase transition at $T_2$ from the paraelectric \textbf{AF3} phase to the ferroelectric \textbf{AF2} one~\cite{Taniguchi06_MWO,Arkenbout_MWO}. The polarization $P_b$ continuously changes through $T_2$ and drops abruptly to zero at $T_1$. The dielectric constant $\epsilon_b$ shows a sharp peak at $T_2$ and a steplike change at $T_1$ when the ferroelectric order disappears~\cite{Taniguchi06_MWO,Arkenbout_MWO}.

The magnetic phase transitions in wolframite were studied theoretically in a number of works~\cite{Harris_Landau_analysis,Toledano_MnWO4,SakhnenkoMnWO4,Matityahu_MWO,Quirion_MWO_CuO}. In~\cite{Harris_Landau_analysis,Toledano_MnWO4} the authors developed phenomenological models of phase transitions in MnWO$_4$ starting with the order parameters belonging to the incommensurate wave vector $\vec{k}_{\rm inc}$. In~\cite{Matityahu_MWO} only the spin components along the easy and $b$ axes are considered, which precludes the description of the magnetic field-induced \textbf{HF} phase as discussed below~\cite{Ehrenberg_MWO}.

In a recent work~\cite{Quirion_MWO_CuO} Quirion and Plumer suggested a Landau theory of magnetic phase transitions in monoclinic multiferroics and applied it to MnWO$_4$ and CuO. The magnetic phase diagrams for various magnetic field directions obtained from the model are in remarkable qualitative agreement with the experimental data. However, from our point of view the suggested model is not relevant neither to MnWO$_4$ nor to CuO due to the following. In the model the magnetic structures of both MnWO$_4$ and CuO are described by a complex pseudovector {\bf S}. This implies that for every direction $\alpha=x$, $y$, and $z$ the magnetic structures are described by a two-dimensional order parameter $({\rm Re}S_\alpha,{\rm Im}S_\alpha)$, which makes the whole magnetic representation 6-dimensional. However, the magnetic representations in both MnWO$_4$ and CuO are 12-dimensional~\cite{SakhnenkoMnWO4,Sakhnenko_Praphase,Sakhnenko_Exchange_FTT}. (Both MnWO$_4$ and CuO possess two magnetic ions in the primitive cell and their magnetic structures are described by the stars of wave vectors possessing two arms, which results in 12-dimensional magnetic representations for Mn$^{2+}$ and Cu$^{2+}$ ions, respectively~\cite{Toledano_BOOK}.) Thus, despite excellent qualitative (and to some extent even quantitative) agreement in the topology of the calculated and experimental phase diagrams the model intrinsically cannot reproduce neither the correct magnetic structures nor various macroscopic properties. For example, in the magnetic structure of the magnetic field-induced phase \textbf{HF} of MnWO$_4$ as obtained in~\cite{Quirion_MWO_CuO} the magnetic moments are directed along the two-fold axis of the monoclinic cell, which contradicts the experimental results~\cite{Ehrenberg_MWO} as well as the prediction from our model as discussed below. We also obtain different predictions for the magnetic structure of the newly discovered phase \textbf{AF3} in CuO~\cite{Villarreal_CuO} as discussed in \sref{sec:CuO}. This illustrates how careful one should be when comparing multiparametric models to experiments.

In our previous work~\cite{SakhnenkoMnWO4} we developed a phenomenological model of phase transitions using magnetic order parameters belonging to the $\vec{k}_{\rm c}$ point of the Brillouin zone and accounted for the incommensurate phases by means of Lifshitz invariants. We also suggested that the phase transitions in wolframite be described starting from the orthorhombic praphase~\cite{SakhnenkoMnWO4,Sakhnenko_Praphase}. In this section using the praphase concept and exchange symmetry we discuss the magnetic structures of the magnetic field-induced phases \textbf{HF}, \textbf{IV}, and \textbf{V} and the magnetic field-induced flop of electric polarization (with magnetic field applied along the easy axis).

In~\cite{SakhnenkoMnWO4} we developed a phenomenological model of magnetic phase transitions in wolframite starting from the monoclinic structure $P2/c$. The magnetically ordered phases are described by the order parameters belonging to the star of wave vector $\vec{k}_{\rm c}$. In this point of the Brillouin zone the space group possesses two two-dimensional IR's $G_1$ and $G_2$. The phases \textbf{AF3} and \textbf{AF1} are described by $G_2$ only, whereas $G_1$ additionally condenses in the phase \textbf{AF2}~\cite{Lautenschlager_MnWO4}. (It has to be noted that in~\cite{SakhnenkoMnWO4,Sakhnenko_Praphase,Sakhnenko_Exchange_FTT} we have inadvertently assumed that the phases \textbf{AF3} and \textbf{AF1} are described by $G_1$ only, whereas $G_2$ additionally condenses in the phase \textbf{AF2}. However, this does not change any of the obtained results.)

Therefore, upon lowering the temperature MnWO$_4$ experiences two close magnetic instabilities with respect to different IR's $G_1$ and $G_2$. It was argued that the phase transitions in wolframite can be described using orthorhombic praphase~\cite{SakhnenkoMnWO4,Sakhnenko_Praphase}. In the monoclinic structure with the monoclinic angle $\beta\approx91^\circ$ the atoms are located at positions Mn - $(\frac{1}{2},0.6853,\frac{1}{4})$, W - $(\frac{1}{2},0.3147,\frac{3}{4})$, O$_1$ - $(0.2108,0.1024,0.9419)$, and O$_2$ - $(0.2516,0.3752,0.3931)$~\cite{Lautenschlager_MnWO4}. The displacement of oxygens towards positions of higher symmetry O$_1$ - $(0.2108,0,0)$ and O$_2$ - $(0.2516,\frac{1}{2},\frac{1}{2})$ and setting $\beta=90^\circ$ results in the orthorhombic structure described by the space group $Pmcm$ (D$_{2h}^5$). In the following we define the orthogonal coordinate axes $x$, $y$, and $z$ parallel to the crystal axes $a$, $b$, and $c$ of the orthorhombic praphase, respectively. The phase transition $Pmcm$-$P2/c$ is described by the homogeneous deformation tensor component $U_{xz}$, which should be taken nonzero when building a model of phase transitions in MnWO$_4$.

The modulation wave vector $\vec{k}_{\rm c}$ retains its position in the orthorhombic structure. In this point of the Brillouin zone the space group $Pmcm$ possesses one four-dimensional IR $P_1$. Thus, the IR's $G_1$ and $G_2$ stem from $P_1$, which splits under the action of $U_{xz}$. Therefore, the magnetically ordered states of MnWO$_4$ can be interpreted as induced by a single IR $P_1$ of the orthorhombic praphase~\cite{SakhnenkoMnWO4,Sakhnenko_Praphase}.

The magnetic representation analysis starting from the praphase can be performed as follows. The magnetic moments of the two Mn$^{2+}$ ions in MnWO$_4$ can be expressed in the form
 \[\vec{M}_1=\left(\begin{array}{c}
 M_{1x}^{\vec{k}_1}\\M_{1y}^{\vec{k}_1}\\M_{1z}^{\vec{k}_1}\end{array}\right) e^{i
 \vec{k}_1\vec{t}} + \left(\begin{array}{c}
 M_{1x}^{\vec{k}_2}\\M_{1y}^{\vec{k}_2}\\M_{1z}^{\vec{k}_2}\end{array}\right) e^{i
 \vec{k}_2\vec{t}},
 \]

  \[\vec{M}_2=\left(\begin{array}{c}
 M_{2x}^{\vec{k}_1}\\M_{2y}^{\vec{k}_1}\\M_{2z}^{\vec{k}_1}\end{array}\right) e^{i
 \vec{k}_1\vec{t}} + \left(\begin{array}{c}
 M_{2x}^{\vec{k}_2}\\M_{2y}^{\vec{k}_2}\\M_{2z}^{\vec{k}_2}\end{array}\right) e^{i
 \vec{k}_2\vec{t}},
 \]
where $\vec{t}$ is the lattice vector, $\vec{k}_1=\vec{k}_{\rm c}$, and $\vec{k}_2=-\vec{k}_1$. The four quantities $M_{i\alpha}^{\vec{k}_j}$ for every direction $\alpha$ transform according to IR $P_1$, i.e., $P_1$ enters three times into the magnetic representation of Mn$^{2+}$ ions. Thus, three order parameters $(g_{1\alpha},g_{2\alpha},g_{3\alpha},g_{4\alpha})$ ($\alpha=x,y,z$), which give the magnetic moment components along $\alpha$ and transform according to $P_1$, determine the magnetic structure in wolframite.

\begin{figure}
\begin{indented}
\item[]\includegraphics{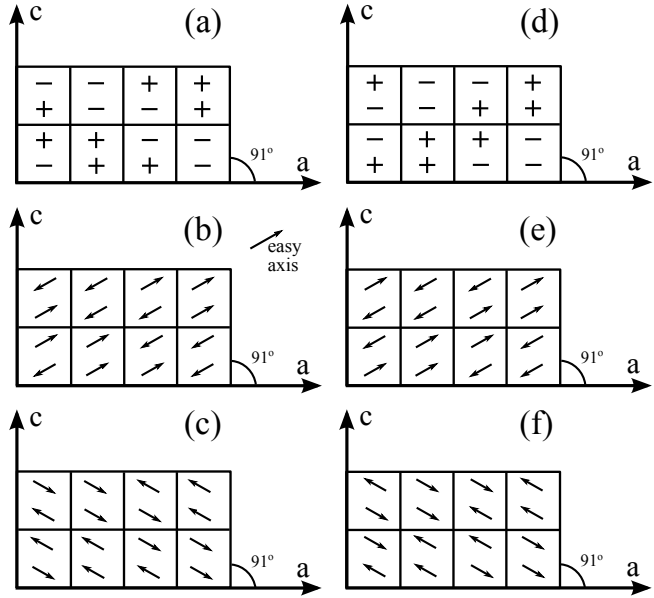}
\end{indented}
\caption{\label{fig:MWO_spins}(a) and (d) give two basic ordering patterns of Mn$^{2+}$ spins corresponding to the exchange multiplets $G_1\otimes V'$ and $G_2\otimes V'$, respectively. The plus and minus signs give relative spin directions. (b) and (c) show magnetic structures in the $ac$ plane with different directions of the easy axis corresponding to the pattern shown in (a), whereas (e) and (f) correspond to (d).
The magnetic structures (b), (c), (e), and (f) of MnWO$_4$ correspond to $q_2\neq0$, $q_4\neq0$, $q_6\neq0$, and $q_8\neq0$, respectively.}
\end{figure}

At low temperatures in the phase \textbf{AF1} (and also in the phase \textbf{HF} as suggested in~\cite{Ehrenberg_MWO}) the magnetic moments are confined to the $ac$ plane. Therefore, in order to determine the magnetic structure of the \textbf{HF} phase we consider the order parameters with $\alpha=x$ and $z$ only. The quadratic part of the thermodynamic potential expansion with respect to these order parameters can be written in the form
\begin{equation}
F=\frac{a}{2}(I_x+I_z)+fI_{xz}+\kappa I_U,\label{eq:MWO_potential}
\end{equation}
where $a$, $f$, and $\kappa$ are phenomenological coefficients, $I_x=g_{1x}^2+g_{2x}^2+g_{3x}^2+g_{4x}^2$, $I_z=g_{1z}^2+g_{2z}^2+g_{3z}^2+g_{4z}^2$, $I_{xz}=g_{1x}g_{1z}+g_{2x}g_{2z}+g_{3x}g_{3z}+g_{4x}g_{4z}$, and $I_U=U'_{xz}(g_{1x}g_{1z}+g_{2x}g_{2z}-g_{3x}g_{3z}-g_{4x}g_{4z})$. The term proportional to $\kappa$ reflects the splitting of the order parameters due to the monoclinic distortion ($U_{xz}$) and external magnetic field $\vec{H}$ applied in the $ac$ plane. Since $H_xH_z$ transforms as $U_{xz}$ under the symmetry elements of the space group and, therefore, directly influences the splitting of the order parameters, we assume $U'_{xz}=U_{xz}-wH_xH_z$, where $w$ is a phenomenological coefficient. In~\eref{eq:MWO_potential} we use the same coefficient at $I_x$ and $I_z$, since the order parameters $g_{ix}$ and $g_{iz}$ belong to the same exchange multiplet~\cite{Sakhnenko_Exchange_FTT}. The diagonalization of the quadratic form~\eref{eq:MWO_potential} results in
\begin{eqnarray}
F&=&(a+f-\kappa U'_{xz})(q_1^2+q_2^2)\nonumber\\
&&+(a-f+\kappa U'_{xz})(q_3^2+q_4^2)\nonumber\\
&&+(a-f-\kappa U'_{xz})(q_5^2+q_6^2)\nonumber\\
&&+(a+f+\kappa U'_{xz})(q_7^2+q_8^2),\label{eq:MWO_potential_Diag}
\end{eqnarray}
where $q_1=(g_{4x}+g_{4z})/2$,  $q_2=(g_{3x}+g_{3z})/2$, $q_3=(g_{4z}-g_{4x})/2$, $q_4=(g_{3z}-g_{3x})/2$, $q_5=(g_{2z}-g_{2x})/2$, $q_6=(g_{1z}-g_{1x})/2$, $q_7=(g_{2x}+g_{2z})/2$, and $q_8=(g_{1x}+g_{1z})/2$. The experimentally observed magnetic structure of the phase \textbf{AF1} is shown in \fref{fig:MWO_spins}(b)~\cite{Ye_MWO}. The easy axis, along which the magnetic moments are aligned in the \textbf{AF1} phase, makes an angle of $35^\circ$ with the $a$ axis. It has to be noted that in~\eref{eq:MWO_potential} we for simplicity did not include the invariants $U'_{xz}(g_{1x}^2+g_{2x}^2-g_{3x}^2-g_{4x}^2)$ and $U'_{xz}(g_{1z}^2+g_{2z}^2-g_{3z}^2-g_{4z}^2)$, which determine the angle between the easy axis and the $a$ axis. The magnetic structure of the \textbf{AF1} phase (i.e., at zero and low fields) is described by the solution $q_2\neq0$. Therefore, from~\eref{eq:MWO_potential_Diag} it can be argued that $f<0$ and $\kappa U'_{xz}>0$, which makes this state preferable at zero applied field and $T<T_1$. (Here we assume that $\kappa>0$ and $U_{xz}>0$ without loss of generality.)

The experimental data reveal that below $T_1$ the application of the magnetic field along the easy axis results in the field-induced \textbf{AF1}-\textbf{AF2}, \textbf{AF2}-\textbf{HF}, \textbf{HF}-\textbf{IV}, \textbf{IV}-\textbf{V}, and \textbf{V}-\textbf{PM} sequence of phase transitions~\cite{Nojiri_MWO,Felea_MWO,Mitamura_MWO}. Information on the magnetic structure of the \textbf{HF} phase is scarce. However, neutron diffraction studies suggest that the field-induced \textbf{HF} phase is commensurate with the same modulation vector as in the phase \textbf{AF1}~\cite{Ehrenberg_MWO,Nojiri_MWO}. It is argued that \textbf{HF} is a spin-flop phase, in which the magnetic moments are switched perpendicular to the applied field within the $ac$ plane with the same relative spin arrangement as in \textbf{AF1}~\cite{Ehrenberg_MWO}. A simple uniform rotation of the spins would result in magnetic structure shown in \fref{fig:MWO_spins}(c), which is given by $q_4\neq0$. In our model we assume that the effective monoclinic splitting $U'_{xz}$ is directly influenced by the magnetic field applied in the $ac$ plane such that $H_xH_z\neq0$. Therefore, it can be argued that sufficiently high magnetic field parallel to the easy axis (i.e., $H_xH_z>0$) changes the sign of the splitting $U'_{xz}$ (thus, $w>0$). Consequently, assuming that $\omega H_xH_z$ is the main contribution to the thermodynamic potential, the solution $q_8\neq0$ will become stable instead of $q_2\neq0$ as follows from~\eref{eq:MWO_potential_Diag}. The corresponding magnetic structure is shown in \fref{fig:MWO_spins}(f) and is different from the one shown in \fref{fig:MWO_spins}(c).

Using the monoclinic lattice as the reference point the magnetic structures of wolframite can be described by two exchange multiplets generated by the IR's $G_1$ and $G_2$~\cite{Sakhnenko_Exchange_FTT}. The respective spin ordering patterns are shown in figures~\ref{fig:MWO_spins}(a) and~\ref{fig:MWO_spins}(d) and can not be transformed into each other by the monoclinic symmetry group operations. The magnetic structure of the phase \textbf{AF1} corresponds to the ordering pattern of \fref{fig:MWO_spins}(a). This pattern is given by the exchange multiplet $G_1\otimes V'$, whose limitation on the space group splits into $G_1\oplus2G_2$, which reflects the fact that both $x$ and $z$ components of magnetic moments are described by IR $G_2$, whereas the $y$ component by $G_1$.

The magnetic structure of the phase \textbf{HF} [\fref{fig:MWO_spins}(c)], which is suggested in~\cite{Ehrenberg_MWO}, corresponds to the same exchange multiplet as that of \textbf{AF1}. However, using the orthorhombic praphase as reference we find that the two multiplets $G_1\otimes V'$ and $G_2\otimes V'$ stem from a single multiplet $P_1\otimes V'$ generated by the IR $P_1$ of the praphase~\cite{Sakhnenko_Exchange_FTT}. The splitting of the exchange energies of the magnetically ordered states corresponding to multiplets $G_1\otimes V'$ and $G_2\otimes V'$ is determined by effective monoclinic distortion $U'_{xz}$, which is directly influenced by the magnetic field with $H_xH_z\neq0$. Therefore, sufficiently high magnetic field along the easy axis changes the sign of $U'_{xz}$, which results in lower exchange energies of magnetically ordered states corresponding to the exchange multiplet $G_2\otimes V'$. Therefore, the magnetic structures of the phases \textbf{AF1} and \textbf{HF} are described by different exchange multiplets ($G_1\otimes V'$ and $G_2\otimes V'$, respectively). Thus, we argue that the phase \textbf{HF} possesses magnetic structure shown in \fref{fig:MWO_spins}(f), which is described by IR $G_1$. Consequently, in the present model the magnetic moments in the phase \textbf{HF} lie in the $ac$ plane perpendicular to the magnetic field direction in accordance with the experimental results~\cite{Ehrenberg_MWO} and in contradiction with the recent model~\cite{Quirion_MWO_CuO}.

The fact that $U'_{xz}$ changes sign at magnetic field $H_{xz}^c\approx 8-10$~T along the easy axis is supported by shrinkage of the temperature range of stability of the phase \textbf{AF3} at $H_{xz}^c$, which results in direct transition from the paramagnetic phase to the phase \textbf{AF2}~\cite{Nojiri_MWO,Felea_MWO}. Therefore, $U'_{xz}=0$ at $H_{xz}^c$, which implies the absence of splitting of IR $P_1$ allowing for simultaneous condensation of IR's $G_1$ and $G_2$ directly from the paramagnetic phase to the phase \textbf{AF2}. At magnetic fields $H_{xz}>H_{xz}^c$ the splitting $U'_{xz}$ has sign different from that at $H_{xz}<H_{xz}^c$, which should favor the condensation of the IR $G_1$ first upon lowering the temperature from the paramagnetic phase. This implies that the magnetic order in the experimentally observed phase \textbf{V}~\cite{Nojiri_MWO,Felea_MWO} is described by IR $G_1$, in contrast to $G_2$ for the phase \textbf{AF3}. Further temperature lowering at $H_{xz}>H_{xz}^c$ should result in additional condensation of $G_2$ and a phase transition to either \textbf{AF2} or \textbf{IV} depending on $H_{xz}$.

According to neutron diffraction experiments, in the phase \textbf{AF3} sinusoidally modulated magnetic moments lie in the $ac$ plane forming an angle with the $a$ axis similar to that in the commensurate \textbf{AF1} phase~\cite{Lautenschlager_MnWO4}. Similarly, the preceding analysis argues that the phase \textbf{V} is incommensurately modulated with magnetic moments lying in the $ac$ plane perpendicular to their direction in the phase \textbf{AF3} and ordered according to the exchange multiplet $G_2\otimes V'$ [\fref{fig:MWO_spins}(d)].

\begin{figure}
\begin{indented}
\item[]\includegraphics{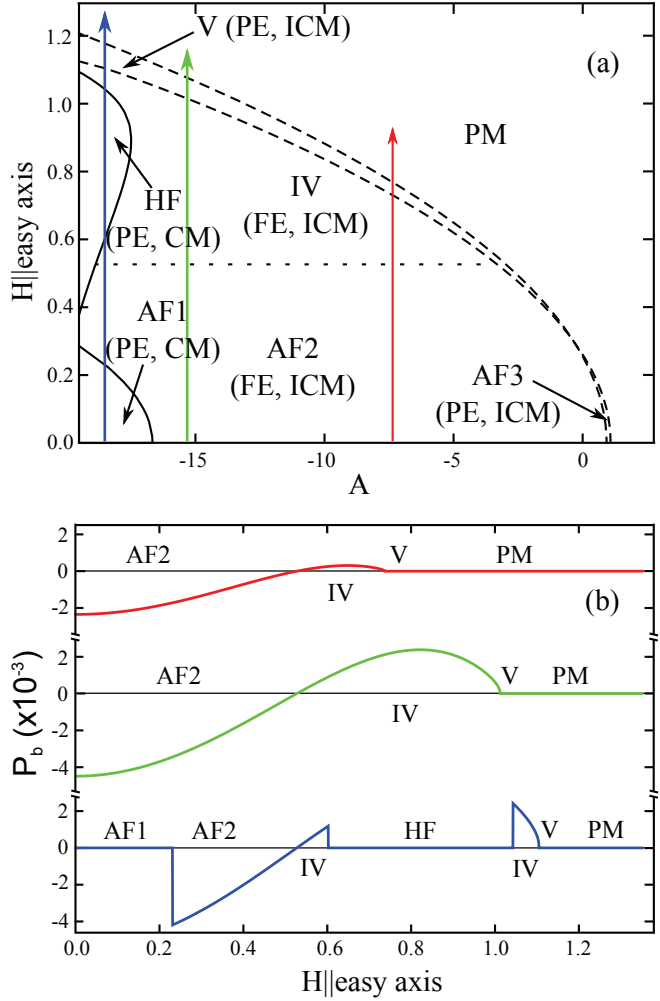}
\end{indented}
\caption{\label{fig:MWO_diag} (colour online) (a) Calculated phase diagram of MnWO$_4$ for magnetic field applied along the easy axis. Dashed and solid lines represent phase transitions of the second and first kind, respectively. Dotted line represents the line at which the electric polarization changes sign. (b) Electric polarization dependence on the magnetic field $H$ at constant $A$. The red, green, and blue lines correspond to the respective paths shown in (a) by coloured axes.}
\end{figure}

To support this interpretation we perform numerical minimization of the following thermodynamic potential expansion with respect to the order parameter $(g_1,g_2,g_3,g_4)$ transforming according to IR $P_1$
\begin{eqnarray}
\Phi & = &\frac{1}{V} \int\left[\frac{A+v(H_x^2+H_z^2)}{2}I_1+\frac{B_1}{4}I_2+\frac{B_2}{4}I_1^2 \right. \nonumber\\
&&\qquad +B_3I_3+ \frac{B_4}{2}I_4+ \frac{B_5}{2}I_5+ \frac{1}{2}I_u + I_p+ \sigma I_L \nonumber\\
&&\qquad\left. +\frac{\delta}{2}I_\delta+\kappa I_p' + \frac{a}{2} P_y^2 \right]dx\label{eq:MnWO4_Potential_Final},
\end{eqnarray}
where the invariants are $I_1=g_1^2+g_2^2+g_3^2+g_4^2$, $I_2=g_1^4+g_2^4+g_3^4+g_4^4$, $I_3=g_1g_2g_3g_4$, $I_4=g_1^2g_2^2+g_3^2g_4^2$, $I_5=g_1^2g_3^2+g_2^2g_4^2$, $I_u=U'_{xz}(g_1^2+g_2^2-g_3^2-g_4^2)$, $I_p=P_y(g_1g_3+g_2g_4)$, $I_p'=H_xH_zU_{xz}P_y(g_1g_3+g_2g_4)$, $I_L=g_2\partial g_1/\partial x - g_1\partial g_2/\partial x+ g_4\partial g_3/\partial x - g_3\partial g_4/\partial x$, and $I_\delta=(\partial g_1/\partial x)^2+(\partial g_2/\partial x)^2+(\partial g_3/\partial x)^2+(\partial g_4/\partial x)^2$, and $V$ is the volume of the sample. In the thermodynamic potential~\eref{eq:MnWO4_Potential_Final}, which is invariant with respect to the space group elements of the praphase, we for simplicity consider the dependence of the order parameter on $x$ only. The following values of phenomenological coefficients $v=15$, $B_1=-21.5$, $B_2=24.75$, $B_3=-40.5$, $B_4=-1$, $B_5=-21$, $w=1$, $\sigma=1$, $\delta=1$, $U_{xz}=0.09$, $\kappa=-40$ and a=500 give the phase diagram shown in \fref{fig:MWO_diag}(a). In the Landau theory of phase transitions it is usually assumed that the coefficient at $I_1$ possesses the strongest dependence on external parameters such as temperature or external magnetic field. Therefore, the $A$ axis in the phase diagram of \fref{fig:MWO_diag}(a) can be associated with temperature. We also included the term proportional to $v$, which describes the simplest magnetic field influence and results in the decrease of temperatures of phase transitions associated with IR $P_1$. It has to be noted that the phenomenological constant $a$ is related to the dielectric susceptibility $\varepsilon$ by $a=\varepsilon^{-1}$.

The invariant $I_u$ reflects the splitting of the order parameter $(g_1,g_2,g_3,g_4)$ into two different order parameters $(g_1,g_2)$ and $(g_3,g_4)$ under the influence of $U'_{xz}$. It can be shown that $(g_1,g_2)$ and $(g_3,g_4)$ transform according to IR's $G_1$ and $G_2$ of the space group $P2/c$, respectively. In turn, the magnetoelectric interaction $I_p$ reflects the necessity of condensation of both IR's $G_1$ and $G_2$ in order for $P_y$ to arise.

The phase diagram shown in \fref{fig:MWO_diag}(a) is obtained by minimizing the functional~\eref{eq:MnWO4_Potential_Final} and is in excellent qualitative agreement with the experimentally observed diagram for magnetic field applied along the easy axis~\cite{Nojiri_MWO,Felea_MWO,Mitamura_MWO}. At zero magnetic field MnWO$_4$ shows the sequence of phase transitions \textbf{AF3}-\textbf{AF2}-\textbf{AF1} at decreasing temperature. Application of magnetic field along the easy axis results in shrinkage of the temperature range of stability of the phase \textbf{AF3}. At $U_{xz}'=0$, which occurs at $H_{xz}^c=0.3$ for the taken values of phenomenological coefficients, the phase \textbf{AF3} disappears corresponding to the absence of effective monoclinic splitting. This value of magnetic field corresponds to the widest temperature range of stability of the phase \textbf{AF2}. Further magnetic field increase results in the growth of $U_{xz}'$ in absolute value but with opposite sign, which leads to the appearance of the phases \textbf{V} and \textbf{HF}, the structures of which were discussed above. The magnetic phases in the phase diagram \fref{fig:MWO_diag}(a) are given by the following values of the order parameter $(g_1,g_2,g_3,g_4)$: \textbf{AF3} - $(0,0,g_3(x),g_4(x))$, \textbf{AF2} and \textbf{IV} - $(g_1(x),g_2(x),g_3(x),g_4(x))$, \textbf{AF1} - $(0,0,g_3,0)$, \textbf{V} - $(g_1(x),g_2(x),0,0)$, and \textbf{HF} - $(g_1,0,0,0)$.

The experiment reveals the field-induced sign change of electric polarization $P_b$ for magnetic fields along the easy axis~\cite{Mitamura_MWO}, which has led to the introduction of the new phase \textbf{IV}~\cite{Nojiri_MWO,Felea_MWO,Mitamura_MWO}. From our point of view such behavior can be explained by the sign change of the coefficient at the magnetoelectric invariant. In~\eref{eq:MnWO4_Potential_Final} we take this into account by the invariant $I_p'$, which together with $I_p$ determines the magnetoelectric response. The resulting magnetic field dependencies of $P_b$ at different values of $A$ are given in \fref{fig:MWO_diag}(b) and are in good qualitative correspondence with the experimental ones~\cite{Mitamura_MWO}.

The experimentally observed wave vector of the incommensurate phase \textbf{IV} $(-0.215,0.503,0.460)$ is close to that of \textbf{AF2}~\cite{Nojiri_MWO}. The appearing modulation along the $b$ axis can be accounted for by the invariant $H_xH_z(g_1\partial g_3/\partial y - g_3\partial g_1/\partial y+ g_2\partial g_4/\partial y - g_4\partial g_2/\partial y)$, which is proportional to the magnetic field with $H_xH_z\neq0$. Therefore, the magnetic field with $H_xH_z\neq0$ also induces spatial modulation of the magnetic order along the $y$ axis due to the Lifshitz invariant proportional to $H_xH_z$.

The experiment reveals the memory effect in MnWO$_4$ for the magnetic field-induced phase transitions sequence \textbf{AF2}-\textbf{HF}-\textbf{IV}, which consists in the fact that irrespective of the poling electric field direction the directions of the electric polarization $P_b$ in the ferroelectric phases \textbf{AF2} and \textbf{IV} are always opposite to each other~\cite{Mitamura_MWO}. This effect persists also when the thermodynamic path crosses the non-ferroelectric phase \textbf{HF}. According to our model the phase \textbf{HF} is paraelectric and the information on the previous polarization direction should be completely lost upon the phase transitions \textbf{AF2}-\textbf{HF} or \textbf{IV}-\textbf{HF}. Therefore, considering the field-increasing run, the subsequent polarization direction in the phase \textbf{IV} when reached from the phase \textbf{HF} should be determined by the applied electric field, which is not the case in the experiment. However, from our point of view this memory effect is dynamic. Indeed, it is found that the characteristic time of the response of the magnetic order parameter in MnWO$_4$ to the electric field pulses is of the order of 10~ms~\cite{Hoffmann_MWO}. This time span is of the order (or even longer) of that of the magnetic field pulses used to study the memory effect~\cite{Mitamura_MWO}. Therefore, the interpretation, which takes into account possible preservation of embryos of the ferroelectric phase~\cite{Mitamura_MWO}, together with the sign change of the magnetoelectric coefficient suggested in our model may explain this memory effect. Experiments with longer magnetic field pulses are required in order to confirm or refute the dynamic nature of the memory effect.

Thus, we have built a model describing the behavior of MnWO$_4$ in magnetic field applied along the easy axis. Using the praphase concept we identified the influence of the magnetic field applied parallel to the easy axis on the thermodynamic potential expansion, which consists not in simple quantitative field dependence of the phenomenological coefficients, but in new terms allowed by the symmetry. This approach also allowed us to suggest possible magnetic structures for the phases \textbf{HF} and \textbf{V}.

It has to be noted, that the phase diagram of \fref{fig:MWO_diag}(a) is topologically similar to that published earlier~\cite{SakhnenkoMnWO4}. In the present work, however, we slightly modified our model and adjusted the phenomenological constants in order to account for new details such as the suppression of the antiferromagnetic order by magnetic field, and magnetically induced flop of electric polarization and the \textbf{AF1}-\textbf{AF2}-\textbf{HF}-\textbf{IV}-\textbf{V} phase transition sequence.

In order to study the problem qualitatively  we for simplicity carried out the expansion of the thermodynamic potential~\eref{eq:MnWO4_Potential_Final} only up to the fourth order with respect to the magnetic order parameter and included only the required magnetic field dependence. The experimental phase diagrams of wolframite are characterized by the values $(T_N-T)/T_N$ close to 1. Therefore, better quantitative comparison with the experiment requires expansion of the thermodynamic potential to higher orders and taking better account of the magnetic field influence, which is beyond the scope of this paper.

\subsection{CuO\label{sec:CuO}}

Cupric oxide CuO is a type-II magnetoelectric with one of the highest Curie temperatures~\cite{KimuraCuO}. At normal conditions it possesses a monoclinic structure described by the space group $C2/c$ (C$_{2h}^6$)~\cite{Yang_CuO}. Upon lowering the temperature CuO undergoes three magnetic phase transitions at $T_{\rm N3}$=230~K, $T_{\rm N2}$=229.3~K, and $T_{\rm N1}$=213~K, which lead to the appearance of magnetically ordered phases \textbf{AF3}, \textbf{AF2}, and \textbf{AF1}, respectively~\cite{Villarreal_CuO}. The phase \textbf{AF2} is characterized by long-wavelength modulation with the wave vector $(0.506,0,-0.483)$, whereas the commensurate phase \textbf{AF1} is described by the wave vector $(\frac{1}{2},0,-\frac{1}{2})$~\cite{KimuraCuO,Ain_CuO}. The existence of the phase \textbf{AF3} was confirmed only recently by ultrasonic measurements and no experimental data on its magnetic structure exist to date~\cite{Villarreal_CuO}. Electric polarization in CuO along the $b$ axis appears in the incommensurate phase \textbf{AF2}~\cite{KimuraCuO}. The newly discovered phase \textbf{AF3} is argued to be paraelectric since no anomaly of the dielectric constant is observed at $T_{\rm N3}$ in contrast to $T_{\rm N2}$~\cite{Villarreal_CuO}.

Magnetic phase transitions and magnetoelectricity in CuO were studied theoretically in a number of works. Monte-Carlo studies based on first-principles calculations failed to reproduce the existence of the phase \textbf{AF3}~\cite{Jin_CuO,Giovannetti_CuO}. A phenomenological theory of phase transitions in CuO was suggested in~\cite{Toledano_CuO}, in which the phase transition \textbf{PM}-\textbf{AF2} was described using the triggering mechanism. This implies the first order character of the \textbf{PM}-\textbf{AF2} transition. In contrast, earlier based on a phenomenological theory we suggested that similar to MnWO$_4$ there should exist an intermediate phase between the paramagnetic and ferroelectric phases~\cite{Sakhnenko_Praphase}. We also suggested an orthorhombic praphase for the description of CuO. Here we further develop this approach to CuO suggesting possible magnetic structure of the phase \textbf{AF3} and building the phase diagram for magnetic fields applied in the $ac$ plane at an angle to both crystal axes, which shows two new field-induced phases.

In the monoclinic structure ($\beta\approx99^\circ$) the copper ions occupy position $(\frac{1}{4},\frac{1}{4},0)$ and oxygen ions - $(0,y,\frac{1}{4})$ ($y=0.416$)~\cite{Yang_CuO}. It can be found that $y=\frac{1}{2}$ and $\beta=90^\circ$ result in the orthorhombic structure described by the space group $Cccm$ (D$_{2h}^{20}$)~\cite{Sakhnenko_Praphase}. Similar to wolframite the phase transition $Cccm$-$C2/c$ is described by the deformation tensor component $U_{xz}$, which transforms according to IR GM$^{4+}$ of the symmetry group of the praphase and has to be assigned nonzero value in order to describe the monoclinic phase.

The low-temperature magnetic structure \textbf{AF1} is described by the wave vector $\vec{k}_c=(\frac{1}{2},0,-\frac{1}{2})$, whereas the incommensurate phase \textbf{AF2} by $(0.506,0,-0.483)$, which is close to the commensurate value~\cite{KimuraCuO,Ain_CuO}. Therefore, one can describe the phase transitions in CuO using the commensurate wave vector $\vec{k}_c$ and account for long-wavelength modulation by Lifshitz invariants allowed by the symmetry~\cite{Sakhnenko_Praphase}. Using the monoclinic structure as reference it was shown that the magnetic structure of the phase \textbf{AF2} is described by two IR's B$_1$ and B$_2$, whereas the phase \textbf{AF1} by B$_2$~\cite{Sakhnenko_Praphase,Sakhnenko_Exchange_FTT,Toledano_CuO}.

The magnetic representation analysis using the orthorhombic praphase can be performed as follows. In the following we define the orthogonal coordinate axes $x$, $y$, and $z$ parallel to the crystal axes $a$, $b$, and $c$ of the orthorhombic structure, respectively. The wave vector $\vec{k}_{\rm c}$ maintains its position in the orthorhombic structure. Thus, the magnetic moments $\vec{M}_1$ and $\vec{M}_2$ of the two copper atoms Cu$_1$ and Cu$_2$ present in the primitive cell and located at positions $(\frac{1}{4},\frac{1}{4},0)$ and $(\frac{1}{4},\frac{3}{4},\frac{1}{2})$, respectively, can be expressed as
 \[\vec{M}_n=\left(\begin{array}{c}
 M_{nx}^{\vec{k}_1}\\M_{ny}^{\vec{k}_1}\\M_{nz}^{\vec{k}_1}\end{array}\right) e^{i
 \vec{k}_1\vec{t}} + \left(\begin{array}{c}
 M_{nx}^{\vec{k}_2}\\M_{ny}^{\vec{k}_2}\\M_{nz}^{\vec{k}_2}\end{array}\right) e^{i
 \vec{k}_2\vec{t}},
 \]
where $n=1$ or 2, $\vec{t}$ is the lattice vector, $\vec{k}_1=\vec{k}_{\rm c}$, and $\vec{k}_2=-\vec{k}_1$. The four quantities $M_{n\alpha}^{\vec{k}_j}$ for every direction $\alpha$ transform according to IR $A_1$ of the space group $Cccm$, i.e., $A_1$ enters three times into the magnetic representation of Cu$^{2+}$ ions. This implies that the two two-dimensional IR's B$_1$ and B$_2$ of the monoclinic space group merge into a single four-dimensional IR $A_1$ of the space group of the praphase. It explains the closeness of instabilities with respect to B$_1$ and B$_2$ on the thermodynamic path and allows interpreting the magnetically ordered states of CuO as induced by a single IR $A_1$~\cite{Sakhnenko_Praphase,Sakhnenko_Exchange_FTT}. Thus, the magnetic representation analysis for CuO closely resembles that for MnWO$_4$ preformed in \sref{sec:MnWO4}. In fact the analogy can be drawn much further.

Let us denote by $(g_1,g_2,g_3,g_4)$ the magnetic order parameter transforming according to $A_1$. It can be shown that $(g_1,g_2)$ and $(g_3,g_4)$ transform according to $B_1$ and $B_2$, respectively, when only the elements of the monoclinic space group are considered. Similar to MnWO$_4$ this is reflected in the existence of the invariant
\[
U'_{xz}(g_1^2+g_2^2-g_3^2-g_4^2),
\]
which splits the order parameter $g_i$~\cite{Sakhnenko_Praphase}. Here we again use $U'_{xz}=U_{xz}-wH_xH_z$ since $H_xH_z$ transforms as $U_{xz}$ under the symmetry elements of the praphase. The magnetoelectric interaction responsible for the appearance of electric polarization $P_y$ in the phase \textbf{AF2} is given by
\begin{equation}
I_p=P_yU''_{xz}(g_1g_3+g_2g_4),\label{eq:CuO_ME_interaction}
\end{equation}
where $U''_{xz}=U_{xz}-w_pH_xH_z$ and $w_p$ is a phenomenological constant. Therefore, condensation of both $B_1$ and $B_2$ is necessarily for its appearance. As follows from~\eref{eq:CuO_ME_interaction} the electric polarization $P_y\propto(U_{xz}-w_pH_xH_z)$ is directly influenced by the magnetic field with $H_xH_z\neq0$. Thus, depending on the sign of $H_xH_z$, which can be changed by choosing suitable field direction, one may expect different magnetic field dependence of $P_y$. For example, if $P_y>0$ then one may either expect its further increase with the magnetic field or, similar to MnWO$_4$, its decrease with subsequent sign change (i.e., a polarization flop).

\begin{figure}
\begin{indented}
\item[]\includegraphics{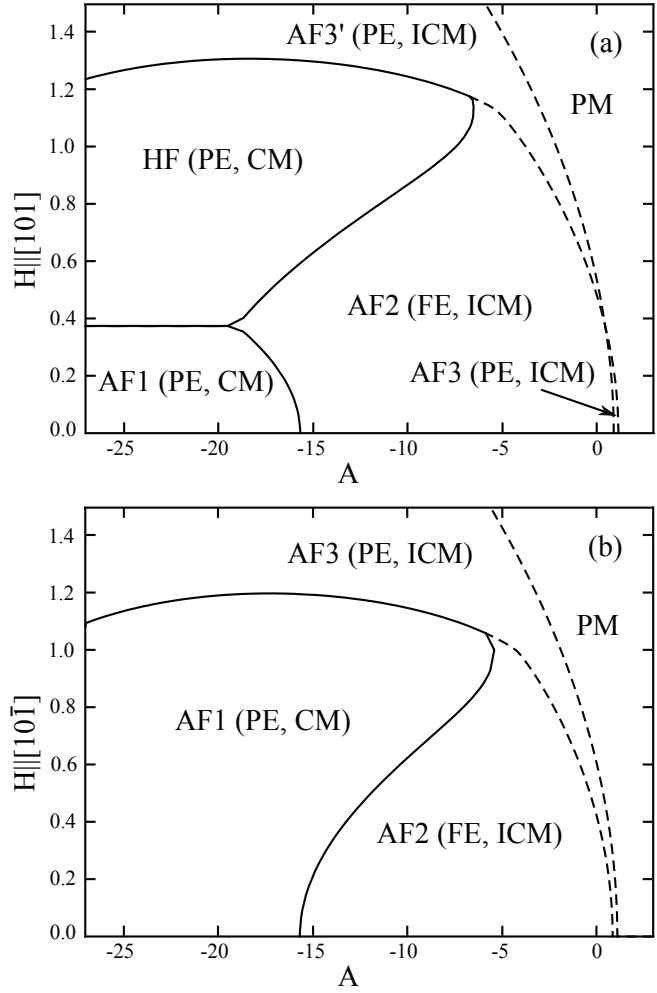}
\end{indented}
\caption{\label{fig:CuO_diag} (a) and (b) give calculated magnetic phase diagrams of CuO for magnetic fields parallel to $[101]$ and $[10\bar{1}]$, i.e., for $H_xH_z>0$ and $H_xH_z<0$, respectively. Dashed and solid lines represent phase transitions of the second and first kind, respectively.}
\end{figure}

The thermodynamic potential expansion with respect to the order parameter $(g_1,g_2,g_3,g_4)$ can be written in the form~\eref{eq:MnWO4_Potential_Final} with the difference that $I_p$ is given by~\eref{eq:CuO_ME_interaction} and $I'_p=0$. The following values of phenomenological coefficients $v=4$, $B_1=-21.5$, $B_2=24.75$, $B_3=-40.5$, $B_4=-1$, $B_5=-21$, $w=1$, $w_p=0$, $\sigma=1$, $\delta=1$, $U_{xz}=0.14$, and a=500 give the phase diagrams shown in \fref{fig:CuO_diag}.
The order parameter takes the following values in the magnetically ordered phases: \textbf{AF3} - $(0,0,g_3(x),g_4(x))$, \textbf{AF2} - $(g_1(x),g_2(x),g_3(x),g_4(x))$, \textbf{AF1} - $(0,0,0,g)$, \textbf{HF} - $(0,g,0,0)$, and \textbf{AF3}$'$ - $(g_1(x),g_2(x),0,0)$. We find that the phases \textbf{AF3}, \textbf{AF2}, and \textbf{AF3}$'$ are incommensurate and are characterized by the same modulation vector, whereas the phase \textbf{AF2} is also ferroelectric with $P_y\neq0$. Similar to wolframite the phases \textbf{AF3} and \textbf{AF1} are described by IR $B_2$, whereas \textbf{HF} and \textbf{AF3}$'$, which corresponds to the phase \textbf{V} of MnWO$_4$, by $B_1$. The phase diagram of \fref{fig:CuO_diag}(a) resembles that of MnWO$_4$ shown in \fref{fig:MWO_diag}(a) due to the similarity in the thermodynamic potential expansion. The external magnetic field along $[101]$ changes the splitting of the order parameter $g_i$, which for sufficiently high fields results in appearance of the phases \textbf{AF3}$'$ and \textbf{HF}, which are described by IR $B_1$. In contrast, the magnetic field parallel to $[10\bar{1}]$ strengthens the splitting resulting in shrinkage of the temperature range of stability of the phase \textbf{AF2}, which is described by both $B_1$ and $B_2$, and widening of the stability ranges of the phases \textbf{AF1} and \textbf{AF3}, which are described by $B_2$. The phase diagrams of \fref{fig:CuO_diag} are obtained under the assumption $w>0$. If $w<0$, then figures~\ref{fig:CuO_diag}(a) and~\ref{fig:CuO_diag}(b) correspond to the magnetic fields applied along $[10\bar{1}]$ and $[101]$, respectively.

\begin{figure}
\begin{indented}
\item[]\includegraphics{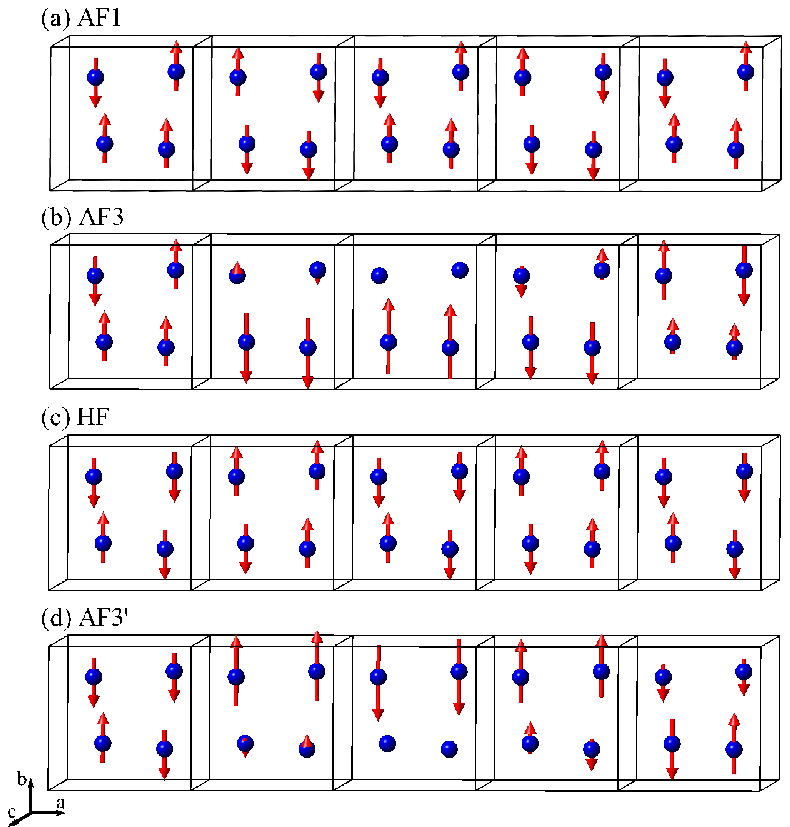}
\end{indented}
\caption{\label{fig:CuO_MagnStructures} (colour online) (a), (b), (c), and (d) give the calculated magnetic structures \textbf{AF1}, \textbf{AF3}, \textbf{HF}, and \textbf{AF3}$'$ of CuO, respectively. The magnetic moments are directed along the $b$ axis.}
\end{figure}
The magnetic representation analysis similar to that performed for MnWO$_4$ in \sref{sec:MnWO4} allows determining the magnetic structures appearing in the phase diagrams of \fref{fig:CuO_diag}. \Fref{fig:CuO_MagnStructures} presents the magnetic structures \textbf{AF1}, \textbf{AF3}, \textbf{HF}, and \textbf{AF3}$'$. In the phase \textbf{AF3} the Cu$^{2+}$ magnetic moments are sinusoidally modulated and directed along the $b$ axis. It has to be noted, that according to our model all of the magnetic moments order in the phase \textbf{AF3} as follows from \fref{fig:CuO_MagnStructures}(b) in contrast to the theoretical result obtained in~\cite{Villarreal_CuO}, where the authors find that only half of the magnetic moments order in the phase \textbf{AF3}. As discussed in \sref{sec:MnWO4} the model of magnetic phase transitions in CuO presented in~\cite{Quirion_MWO_CuO,Villarreal_CuO} suffers from the fact that the magnetic structures of CuO are described by six degrees of freedom, whereas the true magnetic representation is 12-dimensional.

Thus, our model of phase transitions in CuO predicts the magnetic field-induced phase transition \textbf{AF1}-\textbf{HF} for $\vec{H}||[101]$ (or for $\vec{H}||[10\bar{1}]$ if $w<0$). In MnWO$_4$ in a similar phase transition (although through the intermediate phase \textbf{AF2}) the easy axis changes from being parallel to the applied field to perpendicular direction. Therefore, one is tempted to interpret this phase transition as a simple spin-flop phase transition, at which the spins prefer to align along the direction perpendicular to the applied magnetic field. However, as we showed above in \sref{sec:MnWO4} the mechanism of this phase transition can be more complex and deserves additional study of the microscopic causes of the differences in the exchange energies corresponding to two exchange multiplets. In contrast, in CuO as shown in figures~\ref{fig:CuO_MagnStructures}(a) and~\ref{fig:CuO_MagnStructures}(c) this phase transition is between two phases, for both of which the easy axis is the $b$ axis, i.e., perpendicular to the applied magnetic field. Therefore, experimental confirmation of the suggested magnetic field-induced phase transition will strongly support the suggested models of the phase transitions in both MnWO$_4$ and CuO.

\subsection{Pyroxenes NaFeSi$_2$O$_6$ and NaFeGe$_2$O$_6$}\label{sec:Pyroxenes}

Three members of the pyroxene family (with the general formula AMSi$_2$O$_6$) NaFeSi$_2$O$_6$, LiFeSi$_2$O$_6$ and LiCrSi$_2$O$_6$ were recently shown to display ferroelectricity induced by magnetic order~\cite{Jodlauk_Pyroxene}. In this work we focus on NaFeSi$_2$O$_6$ (NFSO) and NaFeGe$_2$O$_6$ (NFGO). In the paramagnetic phase both compounds possess monoclinic symmetry described by the space group $C2/c$ (C$_{2h}^6$)~\cite{Ballet_Pyroxene,Drokina_Pyroxene_NFGO}. Upon cooling the compounds show two magnetic phase transitions at $T_{\rm N}=8$~K and $T_{\rm c}=6$~K for NFSO~\cite{Jodlauk_Pyroxene} and $T_{\rm N}=13$~K and $T_{\rm c}=11.5$~K for NFGO~\cite{Drokina_Pyroxene_NFGO} leading to the magnetically ordered phases \textbf{AF2} and \textbf{AF1}, respectively. Below $T_{\rm N}$ the magnetic structure is found to be incommensurate with $\vec{k}=(0,0.77,0)$~\cite{Mettout_Pyroxene} and $\vec{k}=(0.3357,0,0.0814)$~\cite{Drokina_Pyroxene_NFGO} for NFSO and NFGO, respectively. NFSO is found to be ferroelectric below $T_{\rm c}$ with polarization $P\|b$~\cite{Jodlauk_Pyroxene}, whereas in NFGO the polarization also appears below $T_c$~\cite{Kim_Pyroxene}.

\subsubsection{NaFeSi$_2$O$_6$}

A phenomenological model of phase transitions in NFSO was suggested earlier using the order parameters belonging to the wave vector $(0,0.77,0)$~\cite{Mettout_Pyroxene}. In contrast, we now build a model assuming the instability in the $\vec{k}=(0,\frac{1}{4},0)$ point of the Brillouin zone, which is close to the equivalent vector $(0,0.23,0)$, and account for the spatial modulation by considering the Lifshitz invariants allowed for the respective order parameters. In this point of the Brillouin zone the space group $C2/c$ possesses two two-dimensional IR's $\Lambda_1$ and $\Lambda_2$. In the monoclinic $C2/c$ structure the iron ions Fe$_1$ and Fe$_2$ are located in positions $(0,0.8991,\frac{1}{4})$ and $(0,0.1009,\frac{3}{4})$, respectively. The permutational representation for these ions is given by $2\Lambda_1$. Thus, there are two exchange multiplets given by $\Lambda_1\otimes V'=\Lambda_1\oplus2\Lambda_2$. In the following we define the orthogonal $x$, $y$ and $z$ axes parallel to the $a$ axis, parallel to the $b$ axis and perpendicular to both the $a$ and $b$ axes of the monoclinic cell, respectively. It can be shown that the $y$ components of iron spins transform according to $\Lambda_1$, whereas the $x$ and $z$ components according to $\Lambda_2$. In the magnetically ordered phases the spins are confined to the $ac$ crystal plane~\cite{Ballet_Pyroxene,Mettout_Pyroxene}. Thus, the magnetic order is described by IR $\Lambda_2$. In order to account for the electric polarization $P_b$ one should consider two order parameters $(\eta_x,\xi_x)$ and $(\eta_z,\xi_z)$ transforming according to $\Lambda_2$ and describing the iron spin components along the $x$ and $z$ axes, respectively. The magnetoelectric interaction is then given by $(\eta_x\xi_z-\xi_x\eta_z)P_y$, whereas a single order parameter $(\eta_x,\xi_x)$ or $(\eta_z,\xi_z)$ would require invariants of the form $\eta_\beta \xi_\beta(\eta_\beta^6-7\eta_\beta^4\xi_\beta^2+7\eta_\beta^2\xi_\beta^4-\xi_\beta^6)P_y$, where $\beta=x,z$ and which are of 8th order with respect to the magnetic order parameter. The long-wavelength modulation along the $y$ axis should be accounted for by the Lifshitz invariants $\eta_\beta\partial \xi_\beta/\partial y-\xi_\beta\partial \eta_\beta/\partial y$.

However, the magnetic phase transitions in NFSO allow another interpretation using the praphase concept. Small atomic displacements towards positions of higher symmetry result in the orthorhombic crystal structure with $Cmcm$ (D$_{2h}^{17}$) symmetry~\cite{PSEUDO}. The atomic positions in the monoclinic $C2/c$~\cite{Redhammer_Pyroxene} and orthorhombic $Cmcm$ structures of NFSO are given in \tref{tab:PyroxeneNFSO}.
\Table{\label{tab:PyroxeneNFSO}Atomic positions of NaFeSi$_2$O$_6$ in the monoclinic $C2/c$ and orthorhombic $Cmcm$ structures. In the $Cmcm$ structure the monoclinic angle $\beta=107.3^\circ$ should be set to $\beta=90^\circ$.}
\br
 & \centre{3}{$C2/c$} &\centre{3}{$Cmcm$} \\
\ns
&\crule{3}&\crule{3}\\
   & x & y & z& x & y & z\\
\mr
Na& 0 & 0.3015 & 0.25 & 0 & 0.3015 & 0.25\\
Fe& 0 & 0.8991 & 0.25 & 0 & 0.8991 & 0.25\\
Si& 0.2894 & 0.09 & 0.2343 & 0.2894 & 0.09 & 0.25\\
O$_1$& 0.1139 & 0.0788 & 0.1376 & 0.1139 & 0.0788 & 0.25\\
O$_2$& 0.3599 & 0.2579 & 0.3009 & 0.3599 & 0.2579 & 0.25\\
O$_3$& 0.353 & 0.0085 & 0.0112 & 0.353 & 0 & 0\\
\br
\endtab
It has to be noted that the monoclinic modification of the pyroxene-type MnGeO$_3$ is isostructural to both NFSO and NFGO and possesses significantly smaller monoclinic angle $\beta\approx101.5^\circ$~\cite{Redhammer_Pyroxene_MnGeO3}. The phase transition $Cmcm$-$C2/c$ is described by the component $U_{xz}$ of the homogeneous deformation tensor, which transforms according to IR GM$^{4+}$ of the $Cmcm$ space group. Thus, non-zero $U_{xz}$ should be taken into account when describing the phase transitions in NFSO starting with the orthorhombic praphase.

In the $\vec{k}=(0,\frac{1}{4},0)$ point of the Brillouin zone, which retains in the orthorhombic structure, the space group possesses four two-dimensional IR's $\Delta_i$ ($i=1,2,3,4$). The permutational representation of the iron ions is given by $2\Delta_1$. The limitation of the exchange multiplets, which are given by $\Delta_1\otimes V'$, on the space group splits into the direct sum $\Delta_3\oplus\Delta_2\oplus\Delta_4$, with the $x$, $y$, and $z$ spin components transforming according to $\Delta_3$, $\Delta_2$, and $\Delta_4$, respectively. Thus, one should consider the order parameters $(a_x,b_x)$, $(a_y,b_y)$, and $(a_z,b_z)$ transforming according to three different IR's $\Delta_3$, $\Delta_2$, and $\Delta_4$, respectively. The fact that the order parameters $(\eta_x,\xi_x)$ and $(\eta_z,\xi_z)$ in the monoclinic structure transform according to a single IR is reflected in the existence of the invariant
\[
I_U=(a_xb_z-b_xa_z)U_{xz}.
\]
The magnetoelectric interaction is given by
\[
I_P=(a_xa_z+b_xb_z)U_{xz}P_y,
\]
which is, thus, proportional to the deformation $U_{xz}$.

The expansion of the thermodynamic potential up to the fourth order in powers of the order parameters can be written in the form
\begin{eqnarray}
\Phi&=&\frac{1}{V}\int\left\{\frac{A_1}{2}I_1+\frac{A_2}{2}I_2 + \kappa I_U + f_1I_1^2+f_2I_2^2\right.\nonumber\\
 & & +f_3J_1+f_4J_2 + \sigma_1I_{L1} + \sigma_2I_{L2} + \delta_1I_{\delta1} + \delta_2I_{\delta2} \nonumber\\
&&\left. +sI_P+ \frac{A}{2}P_y^2  \right\}dV,\label{eq:Pyroxene_Potential}
\end{eqnarray}
where $A_1$, $A_2$, $\kappa$, $f_1$, $f_2$, $f_3$, $f_4$,
$\sigma_1$, $\sigma_2$, $\delta_1$, $\delta_2$, $s$, and $A$ are
phenomenological coefficients, $I_1=a_x^2+b_x^2$,
$I_2=a_z^2+b_z^2$, $J_1=(a_x a_z + b_x b_z)^2$, $J_2=(a_x b_z -
b_x a_z)^2$, $I_{L1}=a_x\partial b_x/\partial y - b_x\partial
a_x/\partial y$, $I_{L2}=a_z\partial b_z/\partial y - b_z\partial
a_z/\partial y$, $I_{\delta1}=(\partial a_x/\partial
y)^2+(\partial b_x/\partial y)^2$, and $I_{\delta2}=(\partial
a_z/\partial y)^2+(\partial b_z/\partial y)^2$.

In order to minimize the functional~\eref{eq:Pyroxene_Potential} we introduce polar coordinates
$a_x=r_x\cos \phi$, $b_x=r_x\sin \phi$,
$a_z=r_z\cos(\phi+\Delta\phi)$ and $b_z=r_z\sin(\phi+\Delta\phi)$
and assume that only the phase $\phi$ is spatially dependent.
Minimization of~\eref{eq:Pyroxene_Potential} shows that the
phase transition from the paramagnetic phase to the phase with
$r_x\neq0$, $r_z\neq0$, $P_y=0$, $\Delta\phi=\pi/2$, and
$\phi=-qy$ with
\[
q=\frac{\sigma_1}{2\delta_1}-\frac{(\sigma_1\delta_2-\sigma_2\delta_1)\delta_1^2\kappa^2U_{xz}^2}
{2(A_2\delta_1^2+\sigma_1(\sigma_1\delta_2-2\sigma_2\delta_1))^2}
\]
occurs at
\begin{equation}
A_1=\frac{\sigma_1^2}{\delta_1}+\frac{\delta_1^2\kappa^2U_{xz}^2}
{A_2\delta_1^2+\sigma_1(\sigma_1\delta_2-2\sigma_2\delta_1)},\label{eq:Pyroxene_NFSO_TN}
\end{equation}
where we used the expansion with respect to $\kappa U_{xz}$ up to the
second order. We associate this phase with the phase \textbf{AF2}.

In the ferroelectric phase \textbf{AF1} ($r_x\neq0$, $r_z\neq0$, $P_y\neq0$
and $\Delta\phi\neq\pm\pi/2$) the polarization is given by
$P_y=-sU_{xz}r_xr_z\cos(\Delta\phi)/A$ with
\begin{equation}
\sin\Delta\phi=\frac{A\kappa U_{xz}}{r_xr_z(2Af_3-2Af_4-s^2U_{xz}^2)}.\label{eq:Pyroxene_NFSO_Tc}
\end{equation}
Thus, the phase transition line to this phase is determined by the
condition $|\sin\Delta\phi|=1$, which is satisfied when $|r_xr_z|$
becomes sufficiently large. The temperature range of stability of
the \textbf{AF2} phase is, therefore, proportional to the
distortion $U_{xz}$. Similar to the above cases of MnWO$_4$ and CuO the magnetic field in the $ac$ plane with $H_xH_z\neq0$ will introduce additional contribution to the monoclinic distortion $U_{xz}$, which will affect the invariants $I_U$ and $I_P$ resulting in the field dependence of the transition temperatures $T_{\rm N}$ and $T_{\rm c}$ through~\eref{eq:Pyroxene_NFSO_TN} and~\eref{eq:Pyroxene_NFSO_Tc}, respectively. Thus, for certain field direction and value one may expect zero effective monoclinic distortion, which will result in shrinkage of the phase \textbf{AF2} and a direct transition to \textbf{AF1} from the paramagnetic phase. The external magnetic field applied in the $ac$ plane along $(\mathbf{b}\times\mathbf{c})$ has indeed strong influence on the phase transition temperatures $T_{\rm N}$ and $T_{\rm c}$ and leads to the suppression of $P_b$ and appearance of $P_c$~\cite{Jodlauk_Pyroxene}. In contrast, the magnetic field along $b$ has little impact on the magnetic properties and electric polarization.

The appearance of $P_c$ for $H||(\mathbf{b}\times\mathbf{c})$ can be explained by switching of the spin rotation plane from $ac$ to $bc$ and by the existence of magnetoelectric interactions $(a_yb_z-b_ya_z)P_z$ and $(a_yb_z-b_ya_z)U_{xz}P_x$. The coefficient at the latter in the thermodynamic potential expansion is probably very small since no $P_{(\mathbf{b}\times\mathbf{c})}$ is observed~\cite{Jodlauk_Pyroxene}.

\subsubsection{NaFeGe$_2$O$_6$}

NFGO shows incommensurate magnetic structure at 2.5~K with the modulation vector $(k_x,0,k_z)$, where $k_x=0.323$ and $k_z=0.08$~\cite{Redhammer_Pyroxene}. It is found that upon the appearance of the magnetic structure at $T_{\rm N}$ its modulation vector is given by $k_x\approx0.295$ and $k_z=0.065$. Therefore, on lowering the temperature the $k_x$ modulation vector component grows towards the commensurate value $\frac{1}{3}$~\cite{Redhammer_Pyroxene,Drokina_Pyroxene_NFGO_1}. The commensurate phase, however, is not realized in NFGO. Thus, in order to develop the phenomenological theory of phase transitions in it we use the $\vec{k}=(\frac{1}{3},0,0)$  wave vector. In this Brillouin zone point the space group $C2/c$ has two two-dimensional IR's $B_1$ and $B_2$ and the permutational representation of the ions Fe$_1$ and Fe$_2$ is given by $B_1\oplus B_2$. The magnetic structure in NFGO corresponds to the $B_2\otimes V'$ exchange multiplet, which splits into $2B_1\oplus B_2$.

According to the neutron diffraction experiments the magnetic moments lie predominantly in the $ac$ plane with small component along $b$~\cite{Drokina_Pyroxene_NFGO_1}. Therefore, one can describe the magnetic structure by two order parameters $(\eta_x,\xi_x)$ and $(\eta_z,\xi_z)$, which give the magnetic moments in the $ac$ plane and transform according to $B_1$. It can be argued that the observed small $y$ magnetic moment component also results from the IR $B_1$, which enters, however, into the other exchange multiplet $B_1\otimes V'=B_1\oplus2B_2$. Both order parameters allow Lifshitz invariants $\eta_\beta\partial\xi_\beta/\partial x-\xi_\beta\partial\eta_\beta/\partial x$ and $\eta_\beta\partial\xi_\beta/\partial z-\xi_\beta\partial\eta_\beta/\partial z$ ($\beta=x$, $z$), which are responsible for the long-wavelength modulation in the $ac$ plane. The magnetoelectric interaction is given by $(\eta_x\xi_z-\xi_x\eta_z)P_\beta$. The direction of the electric polarization appearing at $T_{\rm c}$ was not determined experimentally yet, but it can be argued from the aforementioned magnetoelectric interaction that the polarization should lie in the $ac$ plane.

Similar to NFSO the phase transitions in NFGO, which has the same crystal structure, can be described starting from the orthorhombic praphase $Cmcm$. In the $\vec{k}=(\frac{1}{3},0,0)$ point of the Brillouin zone, which retains in the orthorhombic structure, the space group possesses four two-dimensional IR's $\Sigma_i$ ($i=1,2,3,4$). The permutational representation of the Fe$^{3+}$ ions is given by $\Sigma_1\oplus\Sigma_4$ and the magnetic structure is given by the exchange multiplet $\Sigma_4\otimes V'=\Sigma_1\oplus\Sigma_2\oplus\Sigma_3$. Thus, we define the order parameters $(a_x,b_x)$, $(a_y,b_y)$, and $(a_z,b_z)$ transforming according to IR's $\Sigma_3$, $\Sigma_2$, and $\Sigma_1$, respectively. The fact that the order parameters $(\eta_x,\xi_x)$ and $(\eta_z,\xi_z)$ in the monoclinic structure transform according to a single IR is reflected in the existence of the invariant
\[
(a_xa_z+b_xb_z)U_{xz}.
\]
The magnetoelectric interactions are given by
\begin{equation}
(a_xb_z-b_xa_z)P_z,\label{EQ:Pyroxene_NFGO_Pz}
\end{equation}
\begin{equation}
(a_xb_z-b_xa_z)U_{xz}P_x.\label{EQ:Pyroxene_NFGO_Px}
\end{equation}
The description of the phase transition sequence in NFGO using the thermodynamic potential expansion resembles that of NFSO and will not be given here. However, as follows from magnetoelectric invariants~\eref{EQ:Pyroxene_NFGO_Pz} and~\eref{EQ:Pyroxene_NFGO_Px} the $x$ component of the appearing electric polarization $P_x$ is proportional to $U_{xz}$, which may result in its smallness as compared to $P_z$. The discussion on the influence of external magnetic field in the $ac$ plane on the phase transitions and magnetoelectric properties given for NFSO is also valid in the case of NFGO.

\subsection{Cu$_3$Nb$_2$O$_8$\label{sec:Cu3Nb2O8}}

The magnetoelectric effect was recently found in Cu$_3$Nb$_2$O$_8$, which possesses centrosymmetric triclinic structure described by the space group $P\bar{1}$ (C$_i^1$)~\cite{Johnson_Cu3Nb2O8}. Upon lowering the temperature it undergoes two magnetic phase transitions at $T_{\rm N}\approx26$~K and $T_2\approx24$~K leading to two magnetically ordered phases, which we denote \textbf{AF2} and \textbf{AF1}, respectively. The neutron diffraction studies reveal that the magnetic structure in the phase \textbf{AF1} is incommensurate with the crystal lattice and can be described by the wave vector $\vec{k}_m=(0.4876,0.2813,0.2029)$~\cite{Johnson_Cu3Nb2O8}. The electric polarization appears at $T_2$ and reaches values about 17 $\mu$C m$^{-2}$. The magnetic structure of the phase \textbf{AF1} is characterized by Cu$^{2+}$ spins rotating in the plane approximately perpendicular to the reciprocal space direction $(1,2,1)$~\cite{Johnson_Cu3Nb2O8}.

The phenomenological theory of phase transitions can be developed starting from the space group $P\bar{1}$ and using the closest commensurate wave vector $\vec{k}=(\frac{1}{2},\frac{1}{4},\frac{1}{4})$. In this point of the Brillouin zone the space group possesses one two-dimensional IR GP$_1$ and the complete magnetic representation of the Cu$^{2+}$ ions located at positions Cu$_1$ (1a) and Cu$_2$ (2i) is given by 9GP$_1$. In order to describe spins rotating in a plane one has to consider two order parameters $(a_1,b_1)$ and $(a_2,b_2)$, which transform according to the same IR GP$_1$. The spatial modulation of magnetic order is due to the Lifshitz invariants $a_i\partial b_i/\partial\alpha-b_i\partial a_i/\partial\alpha$ ($i=1,2$, $\alpha=x,y,z$), whereas the magnetoelectric interaction is given by
\[
(a_1b_2-b_1a_2)P_\alpha.
\]

The triclinic structure can be transformed to the monoclinic one described by the space group $C2/m$ (C$_{2h}^3$) as shown in \tref{tab:CuNbO}~\cite{PSEUDO}.
\Table{\label{tab:CuNbO}The left panel shows the atomic positions of Cu$_3$Nb$_2$O$_8$ and lattice parameters in the triclinic $P\bar{1}$ structure, whereas the right panel gives those, which result in the monoclinic $C2/m$ structure. The monoclinic structure is characterized by the lattice parameters $a$=9.26 \AA, $b$=6.8 \AA, $c$=5.18 \AA, $\beta$=109.9$^\circ$, and atomic positions Cu$_1$ - (0,0,0), Cu$_2$ - (0.1545,0,0.5403), Nb - (0.5964,0,0.7783), O$_1$ - (0.0511,0.2323,0.7036), O$_2$ - (0.7868,0,0.7365), O$_3$ - (0.3568,0,0.8265).}
\br
&\centre{3}{$P\bar{1}$} & \centre{3}{$C2/m$}\\
& \centre{3}{$a$=5.18 \AA, $b$=5.49 \AA,} & \centre{3}{$a$=5.18 \AA, $b$=5.74 \AA,}\\
& \centre{3}{$c$=6.01 \AA, $\alpha$=72.6$^\circ$,} & \centre{3}{$c$=5.74 \AA, $\alpha$=72.6$^\circ$,}\\
& \centre{3}{$\beta$=83.4$^\circ$, $\gamma$=65.7$^\circ$} & \centre{3}{$\beta$=74$^\circ$, $\gamma$=74$^\circ$}\\
\ns
&\crule{3}&\crule{3}\\
  & x & y & z& x & y & z\\
\mr
Cu$_1$& 0 & 0 & 0 & 0 & 0 & 0\\
Cu$_2$& 0.4597 & 0.0734 & 0.2356 & 0.4597 & 0.1545 & 0.1545 \\
Nb & 0.2217 & 0.5414 & 0.6514 & 0.2217 & 0.5964 & 0.5964 \\
O$_1$& 0.2375 & 0.2032 & 0.8993 & 0.2964 & 0.2833 & 0.8188 \\
O$_2$& 0.2635 & 0.7432 & 0.8304 & 0.2635 & 0.7868 & 0.7868 \\
O$_3$& 0.3554 & 0.7383 & 0.3633 & 0.2964 & 0.8188 & 0.2833 \\
O$_4$& 0.1735 & 0.3072 & 0.4065 & 0.1735 & 0.3568 & 0.3568 \\
\br
\endtab
The phase transition $C2/m$-$P\bar{1}$ is described by the order parameter $U$ transforming according to IR GM$^{2+}$ and representing deformation tensor components $U_{xy}$ and $U_{yz}$.
In the $C2/m$ structure the magnetic order can be described by the order parameters belonging to the $(-\frac{1}{2},0,\frac{1}{2})$ point of the Brillouin zone, in which the space group possesses two two-dimensional IR's B$_1$ and B$_2$. The permutational representation of Cu$_1$ and Cu$_2$ atoms is given by B$_1$ and 2B$_1$, respectively. The multiplets corresponding to IR B$_1$ split into B$_1\otimes V'$=B$_1\oplus$2B$_2$. The $y$ components of spins transform according to B$_1$, whereas the $x$ and $z$ components according to B$_2$ (here the $x$ and $y$ axes of the orthogonal set $(x,y,z)$ are chosen along the $a$ and $b$ crystal axes, respectively). The neutron diffraction experiments~\cite{Johnson_Cu3Nb2O8} allowed determining the direction of the normal to the spin rotation plane in the phase \textbf{AF1}. In the monoclinic structure the normal makes angles of 23$^\circ$, 73$^\circ$, and 124$^\circ$ with the crystal axes $a$, $b$, and $c$, respectively. The precise determination of the would be normal (if the structure were monoclinic) is not possible due to the slight ambiguity in the mutual orientation of the triclinic and monoclinic latices. However it can be argued that the normal lies close to the $ac$ plane. Therefore, in the \textbf{AF1} structure both IR's B$_1$ and B$_2$ condense, which describe the magnetic moment components along $y$ and in the $xz$ plane, respectively.

Similar analysis of the electric polarization direction~\cite{Johnson_Cu3Nb2O8} reveals the angles 13$^\circ$, 82$^\circ$, and 121$^\circ$ with the $a$, $b$, and $c$ axes of the monoclinic cell, respectively. Therefore, all the polarization components $P_a$, $P_b$, and $P_c$ are nonzero with the largest being $P_a$.

We denote by $(p_1,q_1)$ and $(p_2,q_2)$ the magnetic order parameters transforming according to B$_1$ and B$_2$, respectively. The fact that $(a_1,b_1)$ and $(a_2,b_2)$ transform according to the same IR is represented by the invariant
\begin{equation}
(p_1q_2-q_1p_2)U,\label{eq:Cu3Nb2O8_U}
\end{equation}
whereas the magnetoelectric interactions are given by
\begin{equation}
(p_1p_2+q_1q_2)P_b,\label{eq:Cu3Nb2O8_Pb}
\end{equation}
\begin{equation}
(p_1p_2+q_1q_2)UP_a,\quad(p_1p_2+q_1q_2)UP_c.\label{eq:Cu3Nb2O8_PaPc}
\end{equation}
Therefore the first phase transition at $T_{\rm N}$ is connected with the simultaneous condensation of $(p_1,q_1)$ and $(p_2,q_2)$ so that $(p_1q_2-q_1p_2)\neq0$, which results in $(p_1p_2+q_1q_2)=0$ and the absence of electric polarization. In polar coordinates ($p_1=r_1\cos\phi$, $q_1=r_1\sin\phi$, $p_2=r_2\cos(\phi+\Delta\phi)$, $q_2=r_2\sin(\phi+\Delta\phi)$) this corresponds to the phase shift of $\Delta\phi=\pm\pi/2$. The analysis of the magnetic representation shows that the resulting magnetic structure is characterized by the sinusoidal modulation of Cu$_1$ and Cu$_2$ spins with all three spin components $x$, $y$, and $z$ being nonzero. The second phase transition at $T_2$ results in $\Delta\phi\neq\pm\pi/2$ and the appearance of all of the polarization components due to~\eref{eq:Cu3Nb2O8_Pb} and~\eref{eq:Cu3Nb2O8_PaPc}. The polarization direction is close to the crystal axis $a$ as shown above and, therefore, the coefficient at the respective invariant in~\eref{eq:Cu3Nb2O8_PaPc} is the largest. The analysis of the thermodynamic potential is similar to that performed for pyroxenes above (\sref{sec:Pyroxenes}) and will not be given here.

The experimental study of Cu$_3$Nb$_2$O$_8$ in external magnetic fields has not been performed yet. However, according to our model the products of magnetic field components $H_xH_y$ and $H_yH_z$ transform according to GM$^{2+}$ similar to $U$ and allow the interactions $(p_1q_2-q_1p_2)H_xH_y$ and $(p_1q_2-q_1p_2)H_yH_z$, which therefore directly influence the coefficient at $I=(p_1q_2-q_1p_2)$ in the thermodynamic potential expansion. The decrease of the absolute value of the coefficient at $I$ favors shrinking of the phase \textbf{AF2}, whereas its increase results in the growth of the \textbf{AF2} temperature interval and shrinking of the phase \textbf{AF1}.

\subsection{$\alpha$-CaCr$_2$O$_4$\label{sec:CaCr2O4}}

$\alpha$-CaCr$_2$O$_4$ was recently shown to become magnetoelectric below the antiferromagnetic ordering temperature $T_{\rm N}=43$~K~\cite{Chapon_CaCr2O4,Singh_CaCr2O4}. It possesses an orthorhombic crystal structure described by the space group $Pmmn$ (D$_{2h}^{13}$), in which chromium ions form a distorted triangular lattice~\cite{Toth_CaCr2O4}. Below $T_{\rm N}$ an incommensurate magnetic order appears with the wave vector $\vec{k}=(0,0.3317,0)$ and magnetic moments rotating in the $ac$ plane~\cite{Chapon_CaCr2O4,Toth_CaCr2O4}. The electric polarization appears discontinuously below $T_{\rm N}$, though the polycrystalline nature of the sample did not allow the determination of the polarization direction~\cite{Singh_CaCr2O4}.

The neutron diffraction experiments show that the magnetic order is described by two distinct IR's $\Delta_1$ and $\Delta_2$ of the space group~\cite{Chapon_CaCr2O4}. The phenomenological model of phase transitions in $\alpha$-CaCr$_2$O$_4$ was suggested in~\cite{Chapon_CaCr2O4,Singh_CaCr2O4}. Indeed, the magnetic structure can be described by the order parameters $(a_1,b_1)$ and $(a_2,b_2)$ belonging to the wave vector $\vec{k}=(0,\frac{1}{3},0)$ and transforming according to IR's $\Delta_1$ and $\Delta_2$, respectively. In the following we define the $x$, $y$, and $z$ axes parallel to the $a$, $b$, and $c$ crystal axes, respectively. The electric polarization along the $y$ axis is given by $P_y\propto(a_1a_2+b_1b_2)(a_1b_2-b_1a_2)$, i.e., proportional to the fourth order of the magnetic order parameters, which is the possible reason of low values of polarization of the order of 0.5~$\mu$C/m$^2$~\cite{Singh_CaCr2O4}.

Thus, two different IR's $\Delta_1$ and $\Delta_2$ condense at $T_N$ as a result of a weakly first order phase transition~\cite{Chapon_CaCr2O4,Singh_CaCr2O4}. However, such transition is only possible if the coefficients at $a_1^2+b_1^2$ and $a_2^2+b_2^2$ in the thermodynamic potential expansion pass through zero almost simultaneously, which is rarely observed for two distinct IR's. Indeed, one can show that this could be the case. $\alpha$-CaCr$_2$O$_4$ possesses two different sets of chromium ions at positions $(\frac{1}{2},\frac{1}{2},\frac{1}{2})$ and $(0.4932,\frac{1}{4},0.0046)$~\cite{Toth_CaCr2O4}. The permutational representation for both of them in the $(0,\frac{1}{3},0)$ point of the Brillouin zone is given by $\Delta_1\oplus\Delta_2\oplus\Delta_3\oplus\Delta_4$. For both sets of ions the observed magnetic structure is described by the $\Delta_4\otimes V'$ exchange multiplet, which expands into $\Delta_4\otimes V'=\Delta_1\oplus\Delta_2\oplus\Delta_3$. The IR's $\Delta_1$, $\Delta_2$, and $\Delta_3$ describe the components of magnetic moments along the $z$, $x$, and $y$ axes, respectively. Thus, $\Delta_1$ and $\Delta_2$ belong to the same exchange multiplet, which explains the closeness of the above mentioned coefficients.

Further analysis can be performed by noting that the orthorhombic structure of $\alpha$-CaCr$_2$O$_4$ can be represented as a slightly distorted hexagonal lattice. The lattice parameters and the atomic positions of the orthorhombic structure are given in \tref{tab:CaCr2O4}.
\Table{\label{tab:CaCr2O4}Atomic positions of $\alpha$-CaCr$_2$O$_4$ in the orthorhombic $Pmmn$ ($a=11.0579$~\AA, $b=5.8239$~\AA, and $c=5.0553$~\AA) and hexagonal $P6_3/mmc$ [$a=11.0579$~\AA, $b=5.8306$~\AA, and $c=5.0495$~\AA~ (orthorhombic setting)] structures. In the table the atomic positions in the hexagonal structure are given with respect to the orthorhombic lattice. In the hexagonal setting ($a=2.9153$~\AA, $c=11.0579$~\AA) the atoms are at Ca$_1$ - $(1/3,2/3,1/4)$ (occ.=0.25), Ca$_2$ - $(0,0,1/4)$ (occ.=0.25), Cr - $(0,0,0)$ (occ.=1.00), and O - $(2/3,1/3,0.1)$ (occ.=1.00).}
\br
 & \centre{4}{$Pmmn$} & \centre{4}{$P6_3/mmc$}\\
\ns
&\crule{4}&\crule{4}\\
   &  x & y & z& occ. & x & y & z & occ.\\
\mr
Ca$_1$ & 3/4 & 3/4 & 0.3512 & 1.00 & 3/4 & 3/4 & 1/3 & 0.25\\
& & & & & 3/4 & 1/4 & 1/3 & 0.25\\
& & & & & 1/4 & 1/2 & 1/6 & 0.25\\

Ca$_2$ & 1/4 & 3/4 & 0.0385 & 1.00 & 1/4 & 3/4 & 0 & 0.25\\
& & & & & 3/4 & 3/4 & 0 & 0.25\\
& & & & & 1/4 & 0 & 1/2 & 0.25\\

Cr$_1$ & 1/2 & 1/2 & 1/2 & 1.00 & 1/2 & 1/2 & 1/2 & 1.00\\

Cr$_2$ & 0.4932 & 1/4 & 0.0046 & 1.00 & 1/2 & 1/4 & 0 & 1.00\\

O$_1$ & 0.4022 & 1/4 & 0.3365 & 1.00 & 0.4 & 1/4 & 1/3 & 1.00\\

O$_2$ & 0.5904 & 1/4 & 0.6825 & 1.00 & 0.6 & 1/4 & 2/3 & 1.00\\

O$_3$ & 0.5989 & 0.4996 & 0.1665 & 1.00 & 0.6 & 1/2 & 1/6 & 1.00\\
\br
\endtab
The structurally similar compound $\beta$-SrRh$_2$O$_4$ crystallizes in the hexagonal structure $P6_3/mmc$ ($D_{6h}^4$) and has strontium ions disordered over two positions 2b and 2c~\cite{Hector_SrRh2O4}. Similar praphase can be introduced for $\alpha$-CaCr$_2$O$_4$, which has a very weakly distorted hexagonal structure as shown in \tref{tab:CaCr2O4}. Thus, small ion displacements towards positions of higher symmetry and the introduction of Ca disorder similar to that of Sr in $\beta$-SrRh$_2$O$_4$ lead to the $P6_3/mmc$ praphase. In the hexagonal structure the six-fold axis is parallel to the orthorhombic $a$ axis and we define the orthogonal $X$, $Y$, and $Z$ axes parallel to the $-z$, $y$, and $x$ axes respectively. It can be shown that the order-disorder phase transition $P6_3/mmc$-$Pmmn$ is described by the order parameter $(u_1,u_2,u_3,u_4,u_5,u_6)$ [phase state $(0,u,0,0,0,0)$] transforming according to the IR $\Lambda_4$ of the $P6_3/mmc$ space group and belonging to the $(\frac{1}{4},\frac{1}{4},0)$ point of the Brillouin zone.

The two inequivalent sets of four chromium ions in the orthorhombic cell stem from the position 2a of the hexagonal lattice, which is supported by equal magnetic moment magnitudes found for these two sets by neutron diffraction~\cite{Toth_CaCr2O4}. In the hexagonal praphase the appearing magnetic order is described by the star of wave vector $(\frac{1}{6},\frac{1}{6},0)$ ($\Lambda$ point of the Brillouin zone). In this point the permutational representation of chromium ions is given by $\Lambda_1\oplus\Lambda_2$, whereas the observed magnetic structure is described by the $\Lambda_2\otimes V'$ multiplet, which expands into $\Lambda_1\oplus\Lambda_3\oplus\Lambda_4$. The IR $\Lambda_3$ describes the component of magnetic moments parallel to the six-fold axis, whereas $\Lambda_1$ and $\Lambda_4$ the components in the hexagonal plane $XY$. The observed magnetic structure is given by $\Lambda_3$ and $\Lambda_4$. Therefore, the introduction of the praphase does not lead to merging of IR's and the magnetic order is still described by two distinct IR's. Nevertheless, in the praphase two sets of chromium ions present in the orthorhombic structure merge to give only two ions per unit hexagonal cell (compared to eight ions in the $Pmmn$ structure).

The hexagonal praphase allows establishing the hierarchy of interactions in $\alpha$-CaCr$_2$O$_4$. Let us denote by $\eta_i$, $\xi_i$, and $\zeta_i$ ($i=1 - 6$) the magnetic order parameters transforming according to IR's $\Lambda_3$, $\Lambda_4$, and $\Lambda_1$, respectively. The components 2 and 5 of these order parameters belong to the wave vectors, which correspond to $(0,\pm1/3,0)$ in the orthorhombic lattice. The appearing wave vector $(0,\frac{1}{3},0)$ (in the orthorhombic lattice) allows for interactions linear in magnetic field and of third order with respect to the magnetic order parameters such as, for example, $(a_1^3-3a_1b_1^2)H_x$ or $(a_1^2b_2+2a_1b_1a_2-b_1^2b_2)H_z$. However, one can show that when considered from the praphase they become proportional to $u^2$. The first of these interactions stems, for example, from the invariant $H_Z((u_2^2-u_5^2)(\xi_2^3-3\xi_2\xi_5^2)+ (u_1^2-u_4^2)(\xi_1^3-3\xi_1\xi_4^2)+ (u_3^2-u_6^2)(\xi_3^3-3\xi_3\xi_6^2)$, where $u_2=u$ and $u_i=0$ for $i\neq2$. These anisotropic interactions could lead to rich behavior in applied magnetic fields, but it can be argued that coefficients at these invariants are small since they are proportional to $u^2$ and the distortion of the hexagonal lattice does not lead, for example, to the splitting of the energies of states corresponding to different IR's ($\Lambda_3$ and $\Lambda_4$) in the exchange multiplet.

The orthorhombic symmetry allows the invariant $H_xH_z(a_1b_2-a_2b_1)$, which stems from the interaction
\begin{eqnarray}
H_XH_Z(\eta_1\xi_4-\eta_4\xi_1+2\eta_2\xi_5-2\eta_5\xi_2 +\eta_3\xi_6-\eta_6\xi_3)\nonumber \\
+\sqrt{3}H_YH_Z(\eta_4\xi_1-\eta_1\xi_4+\eta_3\xi_6-\eta_6\xi_3).\label{eq:CaCr2O4_HxHz}
\end{eqnarray}
This interaction is not proportional to $u^2$ and may explain the experimentally observed suppression of electric polarization in magnetic field in polycrystalline samples~\cite{Singh_CaCr2O4}. Moreover, the invariants following from the praphase reflect the hexagonal pseudosymmetry present in $\alpha$-CaCr$_2$O$_4$, which can be tested by experiments on single crystals.

Thus, it can be stated that the hexagonal praphase symmetry of $\alpha$-CaCr$_2$O$_4$ reduces to orthorhombic due to the ordering of Ca$^{2+}$ ions. This ordering does not lead to the splitting of energies of states corresponding to $\Lambda_3$ and $\Lambda_4$ stemming from the same exchange multiplet. The system possesses a magnetic instability in the $(\frac{1}{6},\frac{1}{6},0)$ point of the Brillouin zone (in the hexagonal lattice), which has six nonequivalent wave vectors. Due to the orthorhombic distortion the appearing wave structure is described by two wave vectors only, which correspond to the $(0,\frac{1}{3},0)$ modulation vector in the orthorhombic lattice. However, one can expect other arms of the six-vector star to appear under applied magnetic fields due to the interaction~\eref{eq:CaCr2O4_HxHz}, for example, or other invariants such as $H_Z(\xi_1\xi_2\xi_3+\xi_3\xi_4\xi_5-\xi_2\xi_4\xi_6+\xi_1\xi_5\xi_6)$. The two other pairs correspond to the $(0,\frac{1}{3},\frac{1}{2})$ Brillouin zone point of the orthorhombic lattice.

\subsection{FeTe$_2$O$_5$Br}

FeTe$_2$O$_5$Br (FTOB) presents an interesting case of magnetoelectric material~\cite{Zaharko_FeTe2O5Br}. FTOB possesses a monoclinic structure described by the space group $P2_1/c$ (C$_{2h}^5$). Upon lowering the temperature it experiences two magnetic phase transitions at $T_{\rm N1}$=11~K and $T_{\rm N2}$=10.5~K, which result in HT and LT incommensurate structures, respectively. Both of the magnetically ordered phases are described by the modulation vector $(\frac{1}{2},0.463,0)$~\cite{Zaharko_FeTe2O5Br,Pregelj_FeTe2O5Br}. Ferroelectric polarization appears in the $ac$ plane below $T_{\rm N2}$ with the largest component along the $c$ axis ($P_c$=8.5~$\mu$C/m$^2$, $P_a$=1~$\mu$C/m$^2$)~\cite{Pregelj_FeTe2O5Br_1}.

The modulation vector of the magnetically ordered phases is close to the commensurate value $\vec{k}=(\frac{1}{2},\frac{1}{2},0)$, which we use for the description of phase transitions in FTOB. In this Brillouin zone point the space group $P2_1/c$ possesses a single two-dimensional IR $C_1$ and we employ two magnetic order parameters $(a_1,b_1)$ and $(a_2,b_2)$ transforming according to $C_1$. The order parameters allow Lifshitz invariants
\[
I_{L1}=a_1\frac{\partial b_1}{\partial y}- b_1\frac{\partial a_1}{\partial y},\qquad
I_{L2}=a_2\frac{\partial b_2}{\partial y}- b_2\frac{\partial a_2}{\partial y},
\]
which are responsible for the long-wavelength modulation along the $y$ axis.

The magnetoelectric interactions are given by
\begin{equation}
P_y(a_1b_2-b_1a_2)=-P_yr_1r_2\sin(\phi_1-\phi_2),\label{eq:FTOB_MEint_Py}
\end{equation}
\begin{equation}
P_\alpha(a_1b_2+b_1a_2)=P_\alpha r_1r_2\sin(\phi_1+\phi_2),\label{eq:FTOB_MEint_Pxz}
\end{equation}
where $\alpha=x$ or $z$ and we assumed $a_1=r_1\cos\phi_1$, $b_1=r_1\sin\phi_1$, $a_2=r_2\cos\phi_2$, and $b_2=r_2\sin\phi_2$.

The expansion of the thermodynamic potential up to the fourth order in powers of the order parameters can be written in the form
\begin{eqnarray}
\Phi&=&\frac{1}{V}\int \left\{\frac{\alpha_1}{2}I_1+\frac{\alpha_2}{2}I_2+\kappa I' \right.\nonumber\\
 & & +\frac{\beta_1}{4}J_1 + \frac{\gamma_1}{4}I_1^2 + \frac{\beta_2}{4}J_2 + \frac{\gamma_2}{4}I_2^2+fJ\nonumber\\
  & & \left. +\sigma_1 I_{L1} + \sigma_2 I_{L2} + \delta_1 I_{\delta1} + \delta_2 I_{\delta2} \right\}dV,\label{eq:FTOB_potential}
\end{eqnarray}
where $\alpha_1$, $\alpha_2$, $\kappa$, $\beta_1$, $\beta_2$, $\gamma_1$, $\gamma_2$, $f$,
$\sigma_1$, $\sigma_2$, $\delta_1$, and $\delta_2$ are
phenomenological coefficients, $I_1=a_1^2+b_1^2$, $I_2=a_2^2+b_2^2$, $I'=a_1a_2+b_1b_2$ $J_1=a_1^4+b_1^4$, $J_2=a_2^4+b_2^4$, $J=(a_1 b_2 + b_1 a_2)^2$, $I_{\delta1}=(\partial a_1/\partial y)^2+(\partial b_1/\partial y)^2$, $I_{\delta2}=(\partial a_2/\partial y)^2+(\partial b_2/\partial y)^2$.
The analysis of the thermodynamic potential~\eref{eq:FTOB_potential} shows that upon lowering the temperature the paramagnetic phase becomes unstable with respect to simultaneous appearance of spatially modulated $(a_1,b_1)$ and $(a_2,b_2)$ with $r_1\neq0$, $r_2\neq0$, $\phi_1=k_1y$ and $\phi_2=k_2y+\Delta\phi$, where $k_1=k_2$. The appearing magnetic structure is characterized by $\Delta\phi=0$ and can be associated with the HT phase. The magnetoelectric interactions~\eref{eq:FTOB_MEint_Py} and~\eref{eq:FTOB_MEint_Pxz} imply that this phase is paraelectric since~\eref{eq:FTOB_MEint_Py} is identically zero, whereas~\eref{eq:FTOB_MEint_Pxz} averages out to zero upon integration. Further temperature decrease may result in the phase transition to the phase with $\Delta\phi\neq0$ and the appearance of $P_y$. However, the experimental data reveals that the LT phase is ferroelectric with $P_x$ and $P_z$ nonzero~\cite{Pregelj_FeTe2O5Br_1}. The interaction~\eref{eq:FTOB_MEint_Pxz} implies that such electric polarization is possible if $k_1=-k_2$ and $\Delta\phi\neq0$. This magnetically ordered phase can be indeed realized if $\sigma_1$ and $\sigma_2$ possess different signs. Taking into account that $(a_1,b_1)$ and $(a_2,b_2)$ belong to the same exchange multiplet we assume for simplicity $\alpha_1=\alpha_2=\alpha$ and $\sigma_1=-\sigma_2=\sigma$.  Thus, the phase transition to the nonferroelectric HT phase $r_1\neq0$, $r_2\neq0$, and $\phi_1=\phi_2=ky$ occurs at
\[
\alpha=\alpha_{c1}=\frac{\sigma^2}{2\delta}+\frac{\delta\kappa^2}{2\sigma^2}
\]
with
\[
k=\frac{\sqrt{\sigma^2-\alpha\delta}}{\sqrt{2}\delta}.
\]
If $f<0$, subsequent temperature decrease results in the phase transition to the ferroelectric LT phase with $r_1\neq0$, $r_2\neq0$, $\phi_1=ky$, and $\phi_2=-ky+\pi/2$, in which the electric polarization lies in the $ac$ plane according to~\eref{eq:FTOB_MEint_Pxz} as observed experimentally. In the limit of small $\kappa$ this phase transition occurs at
\begin{eqnarray}
\alpha&=&\alpha_{c1}-\nonumber\\
& &\frac{2\delta\Delta_2^2\kappa^2\sigma^2(\Delta_1+\Delta_2-16f)}
{2\Delta_2\sigma^4(\Delta_2-8f)^2-\delta^2\kappa^2(\Delta_1\Delta_2-64f^2)(\Delta_2-4f)},\nonumber
\end{eqnarray}
where $\Delta_1=3\beta_1+4\gamma_1$ and $\Delta_2=3\beta_2+4\gamma_2$. The LT phase is characterized by $k=-\sigma/(2\delta)$. Therefore, the modulation vector experiences a jump at the HT-LT phase transition of the order of $|\delta\kappa^2/(4\sigma^3)|$ in the limit of small $\kappa$. The HT-LT phase transition is, thus, a first order transition, which is experimentally confirmed by the abrupt change of the magnetic structure at $T_{\rm N2}$~\cite{Zaharko_FeTe2O5Br}.

\section{Discussion\label{sec:Discussion}}

In \sref{sec:PT_in_MEs} we have performed the representation analysis of several magnetoelectrics and compared it with the available experimental data. In all the considered cases the electric polarization is induced by two magnetic order parameters. In some cases (MnWO$_4$, CuO, and $\alpha$-CaCr$_2$O$_4$) they transform according to different IR's of the crystal symmetry group of the paraphase, whereas in the remaining cases (pyroxenes, Cu$_3$Nb$_2$O$_8$, and FeTe$_2$O$_5$Br) according to a common IR. The analysis of the exchange Hamiltonian symmetry shows that in all the studied cases, as well as in CuCl$_2$~\cite{Sakhnenko_Exchange_FTT} and magnetoelectric delafossites~\cite{Ter-Oganessian_Delafossites}, the relevant IR's describing the magnetic structures at zero applied external magnetic field belong to the same exchange multiplet. Therefore, according to the concept of a single irreducible representation proposed by Landau~\cite{Landau_v5}, in the considered magnetoelectrics the magnetic structures at zero field are described by a single IR though not of the space group of the paraphase, but of the symmetry group of the exchange Hamiltonian. This conclusion is in accordance with the observation that such is the case in majority of magnetic crystals~\cite{Izyumov_JMMM_4}. However, it should be noted that in roughly 90\% of magnetic crystals the magnetically ordered phases are described by a single IR of the space group of the crystal, which is a stronger statement  and is a consequence of considerable splitting of exchange multiplets by anisotropic interactions~\cite{Izyumov_JMMM_4}. Therefore, it can be concluded that it is an experimental fact that in magnetoelectrics the splitting of exchange multiplets by anisotropic interactions is small, which results in the closeness of instabilities with respect to different IR's entering into the exchange multiplets.

The splitting of exchange multiplets is determined by anisotropic interactions, which possess the symmetry of the crystal~\cite{Izyumov_Neutron_diffraction}. Therefore, latent pseudosymmetries present in magnetoelectric crystals may have strong influence on their magnetic properties. Earlier we have shown that the crystal structures of some magnetoelectrics can be represented as slightly distorted structures (praphases) of higher symmetry~\cite{SakhnenkoMnWO4,Sakhnenko_Praphase}. In this work we further developed this approach and used it in the description of MnWO$_4$, CuO, pyroxenes, Cu$_3$Nb$_2$O$_8$, and $\alpha$-CaCr$_2$O$_4$. Introduction of the praphase has several advantages. In the cases of MnWO$_4$ and CuO two IR's of the space group of the paraphase, which describe the magnetic structures, merge into a single IR of the praphase space group. Therefore, the observed magnetic phase transitions in these magnetoelectrics can be interpreted as between the phases induced by a single order parameter. In the cases of pyroxenes and Cu$_3$Nb$_2$O$_8$ two relevant order parameters transform according to different IR's of the praphase space group, whereas they transform according to the same IR when considered from the paraphase. Thus, two different IR's of the praphase merge into a single IR upon a phase transition to the paraphase. Finally, in the case of $\alpha$-CaCr$_2$O$_4$ two relevant IR's being different in the paraphase description remain different when considered from the praphase. However, in this case the description from the hexagonal praphase significantly reduces the number of chromium ions per unit cell.

In all the cases except $\alpha$-CaCr$_2$O$_4$, the reduction of the crystal symmetry from the praphase to the observed paraphase is described by a tensor quantity $U$ (a component of the deformation tensor), which directly influences the interaction of the two magnetic order parameters. Therefore, external electric or magnetic field or elastic stress, which possesses the same symmetry as $U$, introduces additional symmetry allowed contribution to $U$. The influence of such external effects on the phase transitions were studied in detail for MnWO$_4$ and CuO.

In \sref{sec:MnWO4} we further developed the praphase approach to MnWO$_4$, which was initially suggested in~\cite{SakhnenkoMnWO4}. The praphase-paraphase phase transition in MnWO$_4$ is described by the deformation tensor component $U_{xz}$, which determines the splitting of the 4-dimensional magnetic order parameter into two 2-dimensional order parameters of the monoclinic phase. Therefore, external magnetic field applied along the easy axis (i.e. with $H_xH_z\neq0$) possesses the same symmetry as $U_{xz}$ and directly influences the splitting. This approach allowed us determining the magnetic structures of the magnetic field-induced phases \textbf{HF}, \textbf{IV}, and \textbf{V}. In accordance with the experimental results~\cite{Ehrenberg_MWO} we find that the magnetic moments in the phase \textbf{HF} are confined to the $ac$ plane and directed perpendicular to the applied field. This magnetic structure contradicts the results from the recent model of Quirion and Plumer~\cite{Quirion_MWO_CuO}, in which the magnetic moments in the phase \textbf{HF} are directed along the monoclinic $b$ axis. Our results on the magnetic structures of the phases \textbf{AF3} and \textbf{V} are also in contradiction to their model, which from our point of view suffers from the fact that the magnetic structure of MnWO$_4$ is described by six degrees of freedom, whereas the magnetic representation of Mn$^{2+}$ ions is 12-dimensional. Performing numerical minimization of the thermodynamic potential we obtained temperature -- magnetic field phase diagram, which is in good qualitative correspondence with the experiment. The sign change of the electric polarization $P_b$ for magnetic fields along the easy axis is explained by the sign change of the magnetoelectric invariant and the obtained dependence of $P_b$ is also in good qualitative correspondence with the experiments. However, we argue that the nature of the magnetoelectric memory effect for the magnetic field-induced phase transitions sequence \textbf{AF2}-\textbf{HF}-\textbf{IV} is dynamic.

In \sref{sec:CuO} we applied the praphase concept to CuO. Similarly to wolframite, CuO possesses an orthorhombic praphase and a phase transition to the observed monoclinic structure is described by the deformation tensor component $U_{xz}$. Using the praphase as reference the magnetic structures of CuO are also described by a 4-dimensional IR which, splits into two two-dimensional IR's under the action of $U_{xz}$. Analogously to MnWO$_4$ we build the magnetic phase diagram in magnetic fields parallel to $[101]$ and $[10\bar{1}]$, which thus possess the same symmetry as $U_{xz}$. Our approach allows suggesting the magnetic structures of the newly discovered phase \textbf{AF3}, as well as of the magnetic field-induced phases \textbf{AF3}$'$ and \textbf{HF}. The obtained results are in disagreement with the results of the application of the model of Quirion and Plumer to CuO~\cite{Quirion_MWO_CuO,Villarreal_CuO}, which from our point of view suffers from the same shortcoming as its use in the case of MnWO$_4$.

Similar to the cases of MnWO$_4$ and CuO, the praphase-paraphase phase transition in pyroxenes NaFeSi$_2$O$_6$ and NaFeGe$_2$O$_6$ is described by the deformation tensor component $U_{xz}$. Therefore, external magnetic field applied in the $ac$ plane will possess strong influence on magnetic phase transitions in pyroxenes, which is indeed observed experimentally, in contrast to magnetic field along the $b$ axis~\cite{Jodlauk_Pyroxene}.

The praphase-paraphase phase transition in Cu$_3$Nb$_2$O$_8$ is described by the order parameter $U$ possessing the symmetry of the deformation tensor components $U_{xy}$ and $U_{yz}$. Studies of this compound in magnetic fields were not performed yet. However, it can be argued that the magnetic fields with $H_xH_y\neq0$ and $H_yH_z\neq0$ will have the strongest influence on magnetic phase transitions in Cu$_3$Nb$_2$O$_8$ as discussed in \sref{sec:Cu3Nb2O8}.

In the case of $\alpha$-CaCr$_2$O$_4$ the praphase-paraphase phase transition is described by component $u_2$ of a multi-component non-tensor order parameter with $\vec{k}\neq0$. Nevertheless, the presence of hexagonal pseudosymmetry allows drawing important conclusions concerning the behavior of $\alpha$-CaCr$_2$O$_4$ in external magnetic fields. As discussed in \sref{sec:CaCr2O4} some of the phenomenological interactions allowed by orthorhombic symmetry and reflecting the interaction of magnetic field with the magnetic structure in $\alpha$-CaCr$_2$O$_4$ can be argued to be small, because they become proportional to $u_2^2$ when considered from the hexagonal praphase. Moreover, in the hexagonal praphase the magnetic structure is described by a star with six nonequivalent vectors. Due to orthorhombic distortion only two of them appear upon magnetic phase transitions. However, external magnetic fields applied along certain directions induce magnetic structures corresponding to the remaining four arms. Therefore, additional magnetic reflections in the $(0,\frac{1}{3},\frac{1}{2})$ Brillouin zone point of the orthorhombic lattice should be observed. Single-crystal studies of $\alpha$-CaCr$_2$O$_4$ in external magnetic fields are required in order to confirm these predictions.

\section{Conclusions}

In summary, we performed representation analysis of magnetic phase transitions in several magnetoelectrics using the praphase concept and taking into account the symmetry of the exchange Hamiltonian. This approach allowed explaining the closeness of successive magnetic instabilities on the thermodynamic path and clarifying the behavior of these magnetoelectrics in external magnetic fields of certain directions. Phenomenological models of phase transitions for some magnetoelectrics were suggested.

\ack
The authors acknowledge the financial support by the RFBR grants Nos. 11-02-00484-a and 12-02-31229-mol\_a.



\end{document}